\documentclass[aps,plb,twocolumn,showpacs,amsmath,amssymb,superscriptaddress,nofootinbib,floatfix,preprintnumbers,secnumarabic]{revtex4-1}
\pdfoutput=1

\usepackage[usenames,dvipsnames]{color}
\usepackage{multirow}
\usepackage{graphicx}
\usepackage{adjustbox}
\usepackage{booktabs}
\usepackage{appendix}
\usepackage{slashed}
\usepackage{url}
\usepackage{xspace}
\usepackage{braket}
\usepackage{tabularx}
\usepackage[normalem]{ulem} 
\usepackage{siunitx}
\usepackage[ISO]{diffcoeff}
\usepackage[utf8]{inputenc}
\usepackage{xspace}
\usepackage{bm}
\definecolor{darkblue}{rgb}{0,0,0.5}
\definecolor{darkgreen}{rgb}{0.0,0.5,0.2}
\definecolor{darkred}{rgb}{0.6,0,0}
\usepackage[pdfstartview=XYZ,
bookmarks=true,
colorlinks=true,
linkcolor=darkred,
urlcolor=blue,
citecolor=darkgreen,
pdftex,
bookmarks=true,
breaklinks=true,
linktocpage=true, 
hyperindex=true
]{hyperref}
\usepackage[capitalise]{cleveref}
\usepackage{orcidlink}

\usepackage{fullpage}
\usepackage{geometry}      
\AtBeginDocument{
  \heavyrulewidth=.08em
  \lightrulewidth=.05em
  \cmidrulewidth=.03em
  \belowrulesep=.65ex
  \belowbottomsep=0pt
  \aboverulesep=.4ex
  \abovetopsep=0pt
  \cmidrulesep=\doublerulesep
  \cmidrulekern=.5em
  \defaultaddspace= .5em
  \setlength{\tabcolsep}{0.5em}
}

\RequirePackage[normalem]{ulem}

\newcommand{\Ubl}{$U(1)_{B-L}$\xspace}

\newcommand{\Umt}{$U(1)_{L_\mu-L_\tau}$\xspace}
\newcommand{\Uij}{$U(1)_{L_i-L_j}$\xspace}
\newcommand{\gmu}{$(g-2)_\mu$\xspace}

\newcommand{\cevns}{CE$\nu$NS\xspace}

\newcommand{\Twd}{\ensuremath{T_\mathrm{WD}}\xspace}
\newcommand{\tv}[1]{\ensuremath{\boldsymbol{#1}}}

\begin{document}

\preprint{IFT-UAM/CSIC-24-66}

\vspace*{1.7cm}

\title{
How to rule out $(g-2)_\mu$ in $U(1)_{L_\mu-L_\tau}$ with White Dwarf Cooling
}

\author{Patrick Foldenauer\,\orcidlink{0000-0003-4334-4228}}
\email{patrick.foldenauer@csic.es}
\affiliation{Instituto de F\' isica Te\'orica, Universidad Aut\'onoma de Madrid, 28049 Madrid, Spain} 
\author{Jaime Hoefken Zink\,\orcidlink{0000-0002-4086-2030}}
\email{jaime.hoefkenzink2@unibo.it}
\affiliation{Dipartimento di Fisica e Astronomia, Universit\`a di Bologna, via Irnerio 46, 40126 Bologna, Italy} 
\affiliation{INFN, Sezione di Bologna, viale Berti Pichat 6/2, 40127 Bologna, Italy} 

\date{\today}

\begin{abstract}
In recent years, the gauge group $U(1)_{L_\mu-L_\tau}$ has received a lot of attention since it can, in principle, account for the observed excess in the anomalous muon magnetic moment $(g-2)_\mu$, as well as the Hubble tension. Due to unavoidable, loop-induced kinetic mixing with the SM photon and $Z$, the $U(1)_{L_\mu-L_\tau}$ gauge boson $A'$ can contribute to stellar cooling via decays into neutrinos.
In this work,  we perform for the first time an \textit{ab initio} computation of the neutrino emissivities of white dwarf stars due to plasmon decay in a model of gauged $U(1)_{L_\mu-L_\tau}$. A key result is that current observations of the early-stage white dwarf neutrino luminosity at the 30\% level exclude previously allowed regions of the parameter space favoured by a simultaneous explanation of the $(g-2)_\mu$ and $H_0$ anomalies. In this work, we present the relevant white dwarf cooling limits over the entire $A'$ mass range. In particular, we have performed a rigorous computation of the luminosities in the resonant regime, where the $A'$ mass is comparable to the white dwarf plasma frequencies. 
\end{abstract}

\maketitle

\section{Introduction}
\label{sec:introduction}

Our current best theory of physics at the smallest scales is in terms of  the Standard Model (SM) of particle physics. Despite its enormous success, the SM leaves some of the most pressing questions of elementary particle physics unanswered. 
Most prominently, both an explanation of the small masses of neutrinos, and the existence of dark matter (DM) require physics beyond the SM (BSM). A particularly simple and well-motivated extension of the SM is given by a new gauged \Umt symmetry~\cite{Foot:1990mn,He:1990pn,He:1991qd,Foot:1994vd}. 
This not only allows to accommodate neutrino masses~\cite{Heeck:2011wj,Asai:2018ocx,Bauer:2020itv,Majumdar:2020xws,Singh:2022tvz,Arora:2022hza} and DM~\cite{Baek:2008nz,Baek:2015fea,Biswas:2016yan,Biswas:2016yjr,Foldenauer:2018zrz,Okada:2019sbb,Holst:2021lzm}, but can also help to explain the muon \gmu anomaly~\cite{Baek:2001kca,Ma:2001md,Harigaya:2013twa,Altmannshofer:2016brv}, the Hubble tension~\cite{Escudero:2019gzq,Carpio:2021jhu,Araki:2021xdk}
and  the $b\to s \, \mu\mu$ anomaly~\cite{Altmannshofer:2014cfa,Crivellin:2015mga,Altmannshofer:2016jzy,Chen:2017usq,Baek:2017sew}.\footnote{There still persists a discrepancy between the observed value of the branching ratio of the decay $B^+\to K^+\mu\mu$ and its SM prediction, which favours BSM contributions~\cite{Altmannshofer:2023uci}. However, in light of the recent LHCb result on lepton universality~\cite{LHCb:2022qnv,LHCb:2022vje}, a careful reevaluation of an explanation in terms of a \Umt boson seems to be warranted.}

In the absence of elementary kinetic mixing, the associated gauge boson only couples to second- and third-generation leptons. This makes \Umt models generically hard to test at earth-based laboratory experiments, since couplings to conventional matter, i.e.~quarks and electrons, are only induced at the one-loop level via kinetic mixing with the photon. However, its gauge couplings to mu- and tau-flavoured neutrinos render neutrino interactions an excellent way of searching for \Umt bosons~\cite{Amaral:2021rzw}. In particular, for \Umt bosons lighter than the dimuon threshold almost all leading constraints are due to tests of neutrino physics, like neutrino trident production~\cite{Altmannshofer:2014cfa}, neutrino oscillation measurements at Borexino~\cite{Kaneta:2016uyt,Amaral:2020tga}, the number of effective neutrino degrees of freedom $N_\mathrm{eff}$ during big bang nucleosynthesis (BBN)~\cite{Kamada:2015era,Kamada:2018zxi,Escudero:2019gzq}, or neutrino cooling of supernovae \mbox{(SN)~\cite{Croon:2020lrf,Cerdeno:2023kqo,Akita:2023iwq}}.

A theoretically very clean astrophysical environment to study neutrino physics is provided for by early-stage, hot White Dwarfs (WD). 
Their behaviour is well understood and the various existing equations of state (EoS) for modelling WDs reproduce identical results for the same conditions~\cite{Feynman:1949zz,Salpeter:1961zz,Rotondo:2011zz,mathew2017general,Fantoni:2017mfs}.
In the initial stage of the life of WDs, their evolution is governed by neutrino cooling via plasmon decay into neutrinos exiting the WD core~\cite{Winget:2003xf,Kantor:2007kf}. The cooling through plasmon decay could in principle be enhanced by strong magnetic fields or through the addition of new fields that could connect SM neutrinos with the electron-positron loop~\cite{Landstreet:1967zz,chaudhuri1970neutrino,ca00110y,ca01110v,galtsov1972photoneutrino,DeRaad:1976kd,skobelev1976reaction,Yakovlev1981,Kaminker:1992su,Kennett:1999jh,Bhattacharyya:2005tf,Drewes:2021fjx}. An additional \Umt boson, for example, can in principle enhance the plasmon decay into neutrinos via its kinetic mixing with the photon and thus modify the evolution of WDs. This will ultimately lead to a modification of the WD luminosities compared to the SM prediction. Previously, the resulting constraint on the \Umt parameter space has been estimated~\cite{Bauer:2018onh} via an effective field theory (EFT) analysis of modified plasmon decays in WDs~\cite{Dreiner:2013tja}. In this paper, however, we will perform an \textit{ab initio} calculation of the modified WD luminosities due to a \Umt gauge boson correctly taking into account transverse, axial and longitudinal emissivities as well as the fully gauge-invariant kinetic mixing.

\bigskip

The remainder of this article is organised as follows. In \cref{sec:model}, we introduce the theoretical framework of the minimal \Umt model studied in this paper. In~\cref{sec:wd_cooling}, we present the computation of the neutrino luminosities responsible for WD cooling within \Umt.
Finally, we present our results in~\cref{sec:limits} before presenting our conclusions in \cref{sec:conclusions}.

\section{The \Umt Model}
\label{sec:model}

The Lagrangian of the SM exposes some accidental global symmetries like baryon number, $U(1)_B$, and the lepton family numbers, $U(1)_{L_i}$ with $i=e,\mu,\tau$. Remarkably, the combinations \Ubl and \Uij with $i,j=e,\mu,\tau$ can be promoted to anomaly-free gauge symmetries with only the SM field content.\footnote{Cancelling the gauge anomalies of \Ubl requires the addition of three right-handed, SM-singlet neutrinos. The groups \Uij are already anomaly-free without the addition of right-handed neutrinos (if Majorana mass terms for the neutrinos are forbidden~\cite{Bauer:2020itv}).} 
Among these anomaly-free groups \Umt is of special phenomenological interest as it allows for the explanation of several  experimental anomalies. For example, it can accommodate the observed excess in the anomalous magnetic moment of the muon \gmu, as well as the tension arising from  determining the Hubble constant $H_0$ from early-time cosmology via the cosmic microwave background (CMB)~\cite{Planck:2018vyg} contrasted with the value obtained from local measurements via standard candles like type-Ia supernovae and cepheid variable stars~\cite{Riess:2019cxk}.\footnote{As noted in Ref.~\cite{Blinov:2019gcj}, however, explanations of the $H_0$ tension by light vector mediators are not able to account for the less severe tension in the cosmological parameter $\sigma_8$ linked to the  small scale power spectrum of the universe.}

The relevant parts of the Lagrangian of an extra \Umt symmetry can be compactly  written in matrix form as
{\small
\begin{align}
    \mathcal{L} \supset &- \frac{1}{4}\,
    (B_{\alpha\beta}, W^3_{\alpha\beta}, X_{\alpha\beta})
    \begin{pmatrix}
    1 & 0 & \epsilon_B\\
    0 & 1 & \epsilon_W  \\
    \epsilon_B & \epsilon_W  & 1
    \end{pmatrix}
    \begin{pmatrix}
    B^{\alpha\beta}\\
    W^{3\alpha\beta}\\
    X^{\alpha\beta}
    \end{pmatrix}  \notag \\[2pt]
    &+  \frac{1}{2} \, 
    (B_{\alpha},W^3_\alpha, X_{\alpha})\,
    \frac{v^2}{4}
    \begin{pmatrix}
    g'^2 & g'\,g & 0 \\
    g'\,g & g^2 & 0 \\
    0 & 0 & \frac{4\, M_X^2}{v^2}
    \end{pmatrix}
    \begin{pmatrix}
    B^{\alpha}\\
    W^{3\alpha} \\
    X^{\alpha}
    \end{pmatrix} \notag \\[2pt]
    &- (g'\, j_Y^\alpha,\ g\, j_3^\alpha,\ g_{\mu\tau} \,j^\alpha_{\mu\tau}) 
    \begin{pmatrix}
    B_{\alpha}\\
    W^{3}_{\alpha} \\
    X_{\alpha}
    \end{pmatrix}\,.\label{eq:full_lag}
\end{align}}
Here, $X_\alpha$ denotes the new \Umt gauge boson, while $B_{\alpha\beta}$, $W^3_{\alpha\beta}$  and $X_{\alpha\beta}$ are the hypercharge, neutral $SU(2)_L$ and \Umt field strengths, respectively. Furthermore, $g_{\mu\tau}$ is the \Umt gauge coupling, and  $\epsilon_B$ and $\epsilon_W$ are the kinetic mixing parameters with the hypercharge and neutral weak boson, respectively. {Note that the mixing of the \Umt boson with the neutral weak component, $\epsilon_W/2\, W^3_{\alpha\beta} X^{\alpha\beta}$, is generated at the one-loop level from the $SU(2)_L$ lepton doublets running in the loop~\cite{Bauer:2022nwt}.}
The new gauge boson $X_\alpha$ couples to SM leptons through the gauge current
{\small
\begin{align}
j_{\mu\tau}^\alpha&= \bar L_2 \gamma^\alpha L_2 
          + \bar \mu_R\gamma^\alpha \mu_R 
          - \bar L_3 \gamma^\alpha L_3 -\bar\tau_R\gamma^\alpha \tau_R\,.
\label{eq:lilj_currents}
\end{align}}

The lack of any gauge interactions with conventional matter composed of electrons and quarks (and thus hadrons) sets this gauge group apart from other anomaly-free $U(1)$ extensions. 
At the one-loop level, however, the coupling of the leptophilic gauge boson to the leptons induces an interaction with all SM fermions via a kinetic mixing term with the SM photon and $Z$ boson.
At energies $E\sim T_\text{WD}\ll m_{\mu,\tau}$, we find for these mixings (cf.~\cref{app:mixing} for details)
\begin{align}
\epsilon_A &=\frac{e\, g_{\mu\tau}}{6\pi^2}\ \log\left(\frac{m_\mu}{m_\tau}\right)  \ \sim \ - \frac{g_{\mu\tau}}{70} \,, \\
\epsilon_Z &= - \frac{1}{2} \frac{s_W}{c_W} \epsilon_A \,,
\end{align}
where $s_W$ and $c_W$ are the sine and cosine of the Weinberg angle $\theta_W$. These irreducible loop-induced kinetic mixings are finite and effectively lead to loop-suppressed interactions of the \Umt boson with quarks and electrons. 

In the physical mass basis of the dark photon $A'$, we can express the interactions of the new mediator as (cf.~\cref{app:interactions}),
{\small
\begin{align}\label{eq:phys_int}
    \mathcal{L}_\mathrm{int} =  -g_{\mu\tau} \, j^\alpha_{\mu\tau} \, A'_\alpha
    +e\, \epsilon_A \left( j_\mathrm{EM}^\alpha - \frac{1}{2} \tan^2\theta_W\, j^\alpha_Z \right)A'_\alpha\,,
\end{align}}
with the electromagnetic and $Z$ current defined as
\small{
\begin{align}\label{eq:emcurr}
    j_\mathrm{EM}^\alpha &=\sum_f Q^\mathrm{EM}_f\, \bar f   \gamma^\alpha f \,, \\
    j_Z^\alpha &= \sum_f \bar f \, \gamma^\alpha \ \frac{1}{2}\left[(T^3_f -2\, s_W^2\, Q^\mathrm{EM}_f)   - T^3_f\, \gamma^5\right]  \, f \,. \label{eq:zcurr}
\end{align}}
\normalsize
From~\cref{eq:phys_int} we see that the mass eigenstate of the \Umt boson acquires couplings to the SM electromagnetic and $Z$ currents suppressed by the kinetic mixing parameter $\epsilon_A$.

\subsection{Muon anomalous magnetic moment} 
\label{sec:g2}

\begin{figure}
    \centering
    \includegraphics[width=0.3\textwidth]{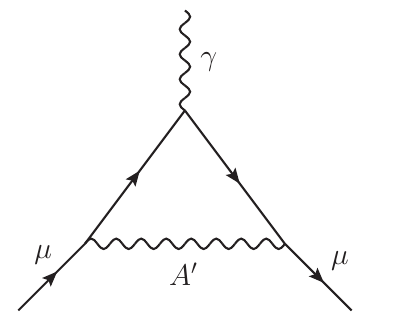}
    \caption{Loop contribution of the \Umt boson to the muon anomalous magnetic moment \gmu.}
    \label{fig:gmu}
\end{figure}

Due to its gauge interactions with the second-generation leptons the \Umt gauge boson $A'$ contributes to the anomalous magnetic moment of the muon, $a_\mu = (g-2)_\mu/2$, via the loop process displayed in~\cref{fig:gmu}. For any neutral gauge boson with vectorial couplings to muons (as in \Umt), the additional contribution to $a_\mu$ can be expressed in the compact form~\cite{Lynch:2001zs,Pospelov:2008zw},
\begin{align}\label{eq:g2mu}
    \Delta a_\mu = Q_\mu^{\prime 2}\ \frac{\alpha'}{\pi} \int_0^1 du \ \frac{(1-u)\, u^2}{u^2+{\displaystyle \frac{(1-u)}{x_\mu^2}} }\,,
\end{align}
where $\alpha'=g_{\mu \tau}^2/4\pi$, $x_\mu=m_\mu/m_{A'}$ and $Q'_\mu$ denotes the \Umt charge of the muon.

In recent years, there has been significant theoretical effort to improve the precision of the SM prediction of \gmu~\cite{Aoyama:2020ynm}, with the current theoretical result being
\begin{align}\label{eq:g2th} 
    a_\mu^\mathrm{SM} = 116\ 591\ 810(43) \times 10^{-11}\,.
\end{align}
At the same time, the E989 experiment at Fermilab has recently reported results from their runs 2 and 3, determining the value of \gmu with unprecedented levels of precision~\cite{Muong-2:2023cdq}. Combined with their run-1 result~\cite{Muong-2:2021ojo} and the previous BNL result~\cite{Muong-2:2006rrc}, the current experimental world average amounts to 
\begin{align}    
    a_\mu^\mathrm{exp} = 116\ 592\ 059(22) \times 10^{-11}\,.
\end{align}
This leads to a $\sim 5.2\sigma$ excess of the experimentally observed value from the theoretical prediction in~\cref{eq:g2th} captured by the total deviation of
\begin{align}    
    \Delta a_\mu = 249(48) \times 10^{-11}\,.
\end{align}
The preferred region in parameter space, where the contribution of a \Umt boson to \gmu can account for this excess is shown by the green band in~\cref{fig:lims}.

It should be noted that a recent lattice result of the leading-order hadronic vacuum polarisation~\cite{Borsanyi:2020mff} significantly decreases the above tension. This, however, comes at the cost of worsening fits to other electroweak precision observables~\cite{Crivellin:2020zul}.

\section{White Dwarf Cooling}
\label{sec:wd_cooling}

\begin{figure}
    \centering
    \includegraphics[width=0.4\textwidth]{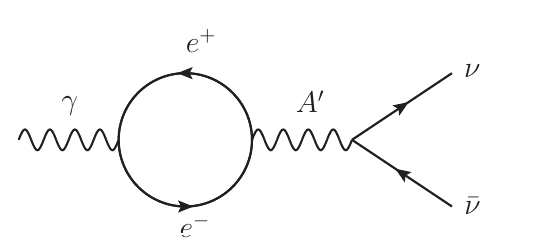}
    \caption{Plasmon decay contribution via \Umt boson coupling to electrons via kinetic mixing.}
    \label{fig:plasmon}
\end{figure}

In this section, we outline the computation of the WD luminosity in neutrinos under the addition of a novel \Umt gauge boson. In doing so we follow closely the computation of Ref.~\cite{Zink:2023szx}. Importantly, we carefully develop the relevant expressions for a light leptophilic gauge boson where plasma effects play a relevant role. We consider a hot white dwarf, in which the main source of energy loss is plasmon decay. We also regard a \Umt leptophilic dark photon that contributes to the plasmon decay. This novel vector mediator has different contributions for each flavour of neutrino-antineutrino pairs that are emitted in the plasmon decay process. In order to obtain more compact expressions, we will introduce the following definitions encapsulating the $A'$ couplings to electrons and neutrinos,
\begin{align}
\label{eq:dark_coeffs}
d_V^e &= e \, \epsilon_A \Big(1 - \tan^2 \theta_W (1 - 4 \sin^2 \theta_W) / 8 \Big)\,, \\
d_A^e &= e\, \epsilon_A\ \tan^2 \theta_W / 8  \,, \\
k_\nu^\alpha &= s_\alpha\, g_{\mu\tau}/2 + d_A^e\,,
\end{align}
where $s_\alpha = 0,1,-1$ for $\alpha=e,\mu,\tau$, respectively. The contribution to the neutrino emissivity of a WD due to a novel \Umt boson originates from the diagram of~\cref{fig:plasmon}, which is identical to the SM neutral current contribution but with a dark photon $A'$ instead of the SM $Z$.
Due to the kinetic mixing coupling of the dark photon to electrons in this model, the BSM contribution can be calculated exactly analogous to the SM $Z$. Hence, we only need to redefine the values of $C_V$ and $C_A$ for each neutrino flavour in Eqs.~(36-38) of Ref.~\cite{Braaten:1993jw} and  consider the full expression for the dark photon propagator to allow for $A'$ masses comparable to the WD plasma frequency,
\begin{equation}
\label{eq:new_coeffs}
\begin{split}
C_a^{\alpha,\text{SM+BSM}}(q) \to C_a^\alpha + b_a\, \frac{\sqrt{2}}{G_F} \frac{k^\alpha_\nu \,d^e_a}{q^2 - m_{A^\prime}^2} \,,
\end{split}
\end{equation}
where $a = V, A$ are the vectorial and axial components with $b_V = 1$ and $b_A = -1$. Here, $q$ is the 4-momentum of the plasmon, $\alpha$ denotes the flavour of the SM neutrino final states, and $C_a^\alpha$ are the coefficients obtained from the SM plasmon decay diagrams: the vectorial ones are equal to $2 \sin^2 \theta_W + 0.5$ (for $e$) and $2 \sin^2 \theta_W - 0.5$ (for $\mu$ and $\tau$), and the axial are $0.5$ (for $e$) and $-0.5$ (for $\mu$ and $\tau$). With these preparations, the WD emissivities $\mathcal{Q}$ into neutrino-antineutrino pairs can be computed following~Ref.~\cite{Braaten:1993jw}. Therefore, we integrate the plasmon decay width $\Gamma_\lambda(q)$ times the energy of the incoming plasmon $\omega_\lambda(\tv q)$ averaged over the photon bosonic thermal distribution $n_B(\omega_\lambda(\tv q), T)$  over the plasmon {3-momentum $\tv q$}. Considering spherical symmetry, the expression for the emissivities read
{\small
\begin{align}
\mathcal{Q}_L = &\frac{G_F^2}{96 \pi^4 \alpha} \!\!\int_0^\infty\!\!\!\!\!\!\! d|\tv{q}| \sum_{\alpha} (C_V^{\alpha,\text{SM+BSM}}(q))^2 \,\tv{q}^2 \, Z_l(\tv{q}) \nonumber\\
&\times\Big(\omega_l(\tv{q})^2 - \tv{q}^2\Big)^2 \,\omega_l(\tv{q})^2 \,n_B (\omega_l(\tv{q}))\,, \label{eq:QL}  \\[8pt]
\mathcal{Q}_T = &\frac{G_F^2}{48 \pi^4 \alpha}\!\! \int_0^\infty \!\!\!\!\!\!\!d|\tv{q}| \sum_{\alpha} (C_V^{\alpha,\text{SM+BSM}}(q))^2 \,\tv{q}^2\, Z_t(\tv{q}) \nonumber\\
&\times\Big(\omega_t(\tv{q})^2 - \tv{q}^2\Big)^3 \,n_B (\omega_t(\tv{q})) \,, \label{eq:QT}\\[8pt]
\mathcal{Q}_A = &\frac{G_F^2}{48 \pi^4 \alpha}\!\! \int_0^\infty \!\!\!\!\!\!\! d|\tv{q}| \sum_{\alpha} (C_A^{\alpha,\text{SM+BSM}}(q))^2\,  \tv{q}^2 \, Z_t(\tv{q}) \label{eq:QA}\\
&\times\Big(\omega_t(\tv{q})^2 - \tv{q}^2\Big) \,  \Pi_A \left(\omega_t(\tv{q}), \tv{q}\right)^2 \, n_B (\omega_t(\tv{q})) \,, \nonumber 
\end{align}}
where  the subscripts $L, T, A$ denote the longitudinal, transverse and axial contributions, respectively.\footnote{Note that the transverse and axial components, $\mathcal{Q}_T$ and $\mathcal{Q}_A$, are both due to transverse polarizations of the plasmon. However, the pure transverse contribution is due to the vectorial coupling $C_V$ to the electron loop and the axial one only to the axial coupling $C_A$. The longitudinal component $\mathcal{Q}_L$ is only containing the vectorial coupling $C_V$.} $Z_l(\tv q)$ and $Z_t(\tv q)$ are the longitudinal and transverse field strength redefinitions of the plasmon and $\Pi_A$ is the axial component of the plasmon self energy.
The 4-momenta of the plasmon have the form $q=\left(\omega_\lambda(\tv{q}), \tv{q}\right)$.

The WD luminosity due to neutrinos is obtained by integrating the emissivities over the volume of the WD core. Considering spherical symmetry, the luminosity is obtained as
\begin{equation}
    L_\mathrm{WD} = 4 \pi \int_{0}^{R_\mathrm{WD}} \mathcal{Q}_{tot}\, r^2 dr\,,
\end{equation}
where $R_\mathrm{WD}$ is the radius of the WD. For our computations, we have chosen a representative value of ${M_\mathrm{WD}=1\, M_\odot}$ for the mass of the WD. However, the results are not altered by picking smaller values. If the star has a temperature of \Twd $\gtrsim 10^{7.8}$ K then the photon luminosity is just $L_\gamma \gtrsim 10^{-0.5} L_\odot$~\cite{shapiro1983physics} and the main cooling mechanism is the plasmon decay inside the star. Therefore, at these temperatures, extra contributions to plasmon decay have an impact on the overall cooling. The limits are computed by estimating the relative excess contribution, $\varepsilon^{\text{BSM}}$, of the novel \Umt boson over the SM one,
\begin{equation}
\label{eq:epsilon}
\varepsilon^{\text{BSM}} \equiv \Big(L_\mathrm{WD}^{\text{SM+BSM}} - L_\mathrm{WD}^{\text{SM}}\Big) / L_\mathrm{WD}^{\text{SM}}\,.
\end{equation}

In the following, we discuss the three regimes of dark photon masses with respect to the WD temperatures:
\begin{enumerate}
    \item[(1)] The \textit{heavy regime} with $m_{A'}\gg$ \Twd.
    \item[(2)] The \textit{ultra-light regime} with $m_{A'}\ll$ \Twd.
    \item[(3)] The \textit{resonant regime} with $m_{A'}\sim$ \Twd.
\end{enumerate}

\subsection{Heavy dark photons} 
\label{sec:heavyDP}

A WD cannot reach temperatures for which the energies of the plasmon decay are substantially greater than $\mathcal{O}(10)$ MeV. Therefore, in the heavy regime of $m_{A'}\gtrsim 10$ MeV the dark photon propagator entering the momentum-integral in the emissivity can be well approximated as ${1 / (q^2 - m_{A^\prime}^2) \sim -1 / m_{A^\prime}^2}$. Hence, $C_V^{\alpha,\text{SM+BSM}}$ and $C_A^{\alpha,\text{SM+BSM}}$ no longer depend on $q$ and we can directly compute $\varepsilon^{\text{BSM}}$ since the axial contribution is negligible,
\begin{equation}
\label{eq:epsilon_heavy}
\begin{split}
\varepsilon^{\text{BSM}} &= \sum_{\alpha} \Big(C_V^{\alpha,\text{SM+BSM}}\Big)^2 / \sum_{\alpha} \Big(C_V^{\alpha,\text{SM}}\Big)^2 - 1 \\
&\simeq 1.50 \times 10^{17} \bigg(\frac{g_{\mu\tau}}{m_{A^\prime}/1~\mathrm{MeV}}\bigg)^4 \\ & \ \ \ \ - 1.66 \times 10^{5} \bigg(\frac{g_{\mu\tau}}{m_{A^\prime}/1~\mathrm{MeV}}\bigg)^2\,.
\end{split}
\end{equation}
This quantity is also independent of the exact temperature of the WD, but is only valid for masses that are much greater than the plasma frequency of the WD\footnote{See \cref{eq:plasma_freq} for the full expression of the plasma frequency.}. 
Due to the relative smallness of $\epsilon_Z$ compared to $\epsilon_A$, the results are also unaltered if we neglect the dark photon coupling to the $Z$ current in~\cref{eq:phys_int}.

\subsection{Ultra-light dark photons} 
\label{sec:ultra-light-A}

In the case of an ultra-light dark photon mediator,
effects due to the self energy correction to the propagator become relevant when computing its propagation in the stellar medium and the subsequent plasmon decay.
The dark photon self energy is given by
\begin{equation}
\begin{split}
\Pi_{A^{\prime}}^{\mu \nu} (q) = &- \int \frac{d^4 k}{(2\pi)^4} \ \mathrm{tr} \big[ \gamma^\mu (d_V^e + d_A^e \gamma^5) (\slashed{k} + m_{e})\\
    &\times \gamma^\nu (d_V^e + d_A^e \gamma^5) (\slashed{q} - \slashed{k} - m_{e}) \big]\\
    &\times \bigg\{ \frac{i}{k^2 - m_{e}^2} - 2\pi\big[ \theta(-k^0) \\
    & + \mathrm{sign}(k^0)\, \tilde{f}(k^0 - \mu) \big] \delta\big(k^2 - m_{e}^2\big) \bigg\}\\
    &\times \bigg\{ \frac{i}{(q-k)^2 - m_{e}^2} - 2\pi\big[ \theta(-q^0 + k^0) \\
    & + \mathrm{sign}(q^0 - k^0)\, \tilde{f}(q^0 - k^0 + \mu) \big] \\
    & \times\delta\big((q-k)^2 - m_{e}^2\big) \bigg\} \,.
\end{split}
\end{equation}
where $\tilde{f}(x) \equiv (e^{\beta x} + 1)^{-1}$ with  $\beta = 1 / (k_B\, T)$.

We can conveniently express this in terms of the plasmon self energy $\Pi_{\gamma}^{\mu \nu}$. To do so, we note that $(d_A^e)^2 / (d_V^e)^2 \sim \mathcal{O}(10^{-3})$ is negligible, as well as the axial contribution to the self-energy, which is $\sim d_V^e \,d_A^e\, \Pi_{A,\gamma}$. This is warranted since the axial contribution to the plasmon self-energy $\Pi_{A,\gamma}$ appearing is typically four to six orders of magnitude smaller than the longitudinal and transverse ones. Taking this into account, the final expression for the dark photon self energy reads
\begin{equation}
\label{eq:Ap_selfE}
\begin{split}
\Pi_{A^{\prime}}^{\mu \nu} (q) = &\frac{(d_V^e)^2}{4\pi \alpha}\, \Pi_{\gamma}^{\mu \nu} (q)\,, 
\end{split}
\end{equation}
with $\Pi_{\gamma}^{\mu \nu}$ the plasmon self-energy. 
Next, we need to compute the full propagator, $D^{\mu\nu}_{A^{\prime}}$, up to the same order, $\mathcal{O}(\alpha)$, used for the plasmon decay computation. To compute this quantity, we need to expand the self-energy into its longitudinal and transverse components. We can work with the same projectors as used for the photon since, as we saw, the dark photon self-energy is proportional to that of the photon. Hence, we can write
\begin{equation}
\Pi_{A^{\prime}}^{\mu \nu} = F_{A^{\prime}} P_L^{\mu \nu} + G_{A^{\prime}} P_T^{\mu \nu} \,, \label{eq:Ap_selfE_TL}
\end{equation}
with the transverse and longitudinal projectors given by
\begin{equation}
\begin{split}\label{eq:projs}
P^{\mu \nu}_T &= \big( \delta^{ij} - \hat{q}^i \hat{q}^j\big)\, \delta^\mu_i \delta^\nu_j \,, \\
P^{\mu \nu}_L &= \bigg( -g^{\mu \nu} + \frac{q^\mu q^\nu}{q^2} \bigg) - P^{\mu \nu}_T \,.
\end{split}
\end{equation}
Here, the factors $F_{A^{\prime}}$ and $G_{A^{\prime}}$ are obtained analogously to the photon case, by contracting $\Pi^{\mu\nu}_{A'}$ with the projectors of~\cref{eq:projs}. Formally, we can write this as
\begin{equation}
F_{A^{\prime}} \equiv \Pi_{A^{\prime}}^{00}\, \frac{q^2}{\tv{q}^2} \,, \qquad G_{A^{\prime}} \equiv \Pi_{A^{\prime}}^{xx}\,,
\end{equation}
where $x$ is any  direction transverse to the propagation of the dark photon with $4-$momentum $q = (\omega, \tv{q})$.

Importantly, we note that since $\Pi_{A^{\prime}}^{\mu \nu}$ is proportional to the plasmon self-energy, its contraction with $q^\mu$ vanishes. Hence it is respecting the Ward identity, even though the gauge symmetry is broken. In our computation, the full propagator $D^{\mu\nu}_{A^\prime}$ is contracted with the plasmon self-energy $\Pi_{\gamma}^{\mu \nu}$. Thus, we will neglect any term proportional to $q^\mu$. 
Furthermore, we remind ourselves that for the projectors {${P}^{\mu}_{T\lambda} \, P^{\lambda \nu}_T = P^{\mu \nu}_T$, ${P}^{\mu}_{L \lambda} \, P^{\lambda \nu}_L = P^{\mu \nu}_L$, and ${P}^{\mu}_{L \lambda} \, P^{\lambda \nu}_T = P^{\mu}_{T\lambda}\,  P^{\lambda \nu}_L = 0$.
With these preparations, we can write the expression for the full propagator of the dark photon as 
\begin{widetext}
\begin{equation}\label{eq:full_prop}
\begin{split}
D_{A^{\prime}}^{\mu \nu} &= \frac{-i\,(g^{\mu \nu} - q^\mu q^\nu/m^2_{A^{\prime}})}{q^2 - m^2_{A^{\prime}}} 
+ \frac{-i\,(g^{\mu}_{\lambda} - q^\mu q_\lambda/m^2_{A^{\prime}})}{q^2 - m^2_{A^{\prime}}} \left(i\,\Pi_{A^{\prime}}^{\lambda \sigma}\right) \frac{-i\,(g_{\sigma}^{\nu} - q_\sigma q^\nu/m^2_{A^{\prime}})}{q^2 - m^2_{A^{\prime}}} + \mathrm{...}\\[8pt]
&=\frac{-i \,g^{\mu \lambda}}{q^2 - m^2_{A^{\prime}}} \ \Bigg[\delta_{\lambda}^{\nu} + \sum_{n=1}^\infty \bigg(\frac{F_{A'}}{q^2 - m^2_{A^{\prime}}} \bigg)^n P_{L\lambda}^{\nu} + \sum_{n=1}^\infty \bigg(\frac{G_{A'}}{q^2 - m^2_{A^{\prime}}} \bigg)^n P_{T\lambda}^{ \nu}\Bigg]\\[8pt]
&=\frac{-i \,g^{\mu \lambda}}{q^2 - m^2_{A^{\prime}} - F_{A'}} \ P_{L\lambda}^{ \nu} + \frac{-i \,g^{\mu \lambda}}{q^2 - m^2_{A^{\prime}} - G_{A'}} \ P_{T\lambda}^{ \nu} \,,
\end{split}
\end{equation}
\end{widetext}
where finally we have
\begin{align}
  F_{A^\prime} &\equiv \frac{(d_V^e)^2}{4\pi \alpha}\ \frac{q^2}{ \tv{q}^2}\  \Pi^{\gamma}_L \,, \\
  G_{A^\prime} &\equiv \frac{(d_V^e)^2}{4\pi \alpha}\  \Pi^{\gamma}_T\,, 
\end{align}
with $\Pi^{\gamma}_L$ and $\Pi^{\gamma}_T$ the plasmon longitudinal and transverse self-energies, that can be found in~\cite{Zink:2023szx} in Eqs.~(18) and (19).
To obtain the last line of~\cref{eq:full_prop}, we have made use of the identity 
\begin{equation}
\delta^{\lambda \nu} = P^{\lambda \nu}_L + P^{\lambda \nu}_T + q^\lambda q^\nu / q^2\,.
\end{equation}

In the computation of the emissivities the full $A'$ propagator $D_{A^{\prime}}^{\mu \nu}$ is contracted with the plasmon self-energy $\Pi^{\mu\nu}_\gamma$. Hence, we find that only the longitudinal component of $D_{A^{\prime}}^{\mu \nu}$ enters the longitudinal emissivity and only its transverse component enters the transverse and axial emissivities. Furthermore, since the $A'$ self-energy can be expressed in terms of the photon self-energy and the $4$-momentum at which it is evaluated is the on-shell plasmon momentum (due to momentum conservation), we can use the standard plasmon relations to evaluate $P^{\mu \nu}_L$ and $P^{\mu \nu}_T$ for the dark photon. Taking this into consideration, the expression for the full propagator finally simplifies to
{\small
\begin{align}
\label{eq:full_prop_dark_photon}
D_{A^{\prime}}^{\mu \nu} = &\frac{-i\, P^{\mu \nu}_L}{q^2 - m^2_{A^{\prime}} - \cfrac{(d_V^e)^2 + (d_A^e)^2}{4\pi \alpha}\ \big(\omega_l(\tv{q})^2 - \tv{q}^2\big)} \notag \\[8pt] 
+&\frac{-i\, P^{\mu \nu}_T}{q^2 - m^2_{A^{\prime}} - \cfrac{(d_V^e)^2 + (d_A^e)^2}{4\pi \alpha}\ \big(\omega_t(\tv{q})^2 - \tv{q}^2\big)}\,.
\end{align}
}
\normalsize
This expression for the full propagator has to be used instead of the naive tree-level propagator $\frac{-i}{{q^2 - m_{A^\prime}^2}}$ in computing the coupling coefficients in \cref{eq:new_coeffs}. Effectively, this means that in the computation of the longitudinal emissivity the denominator of the coefficient $C_V^{\alpha,\text{SM+BSM}}(q)$ has to be replaced by the one of the first term in~\cref{eq:full_prop_dark_photon}. Similarly, for the transverse and axial components the denominator of the coupling coefficients $C_V^{\alpha,\text{SM+BSM}}(q)$ and $C_A^{\alpha,\text{SM+BSM}}(q)$ have to be replaced by the denominator of the second term in~\cref{eq:full_prop_dark_photon}.

In the very low mass region, where $m_{A'}^2\ll q^2$, the luminosities become approximately independent of the mass. To see this, let us consider the denominator of the propagator in~\cref{eq:full_prop_dark_photon}. This has the form $(1 - r_\mathrm{BSM})\, q^2 - m_{A^\prime}^2$, where $r_\mathrm{BSM} \equiv \left[(d_V^e)^2 + (d_A^e)^2\right]/{4\pi \alpha}$ . In this region the denominator of the propagator can be approximated as $\sim 1 / q_r^2$, where $q_r^2 \equiv (1 - r_\mathrm{BSM})\,q^2$. The luminosity in this region is then obtained by integrating the emissivities in \cref{eq:QL,eq:QT,eq:QA} with the replacements,
\begin{widetext}
{\footnotesize
\begin{equation}
\begin{split}
\sum_\alpha (C_V^{\alpha,\text{SM+BSM}}(q))^2 &= \frac{d_e^V}{G_F^2\, (q_r^2)^2} \bigg(d_e^V \big(6\, (d_e^A)^2 + g_{\mu \tau}^2 \big) + \sqrt{2}\, G_F \, q^2_r \Big[ 2 \,d_e^A \,\sum_\alpha C_V^{\alpha,\text{SM}}  + g_{\mu \tau}\, \big(C_V^{\mu,\text{SM}} - C_V^{\tau,\text{SM}} \big)   \Big] \bigg)\,,\\
\sum_\alpha (C_A^{\alpha,\text{SM+BSM}}(q))^2 &= \frac{d_e^A}{G_F^2 (q_r^2)^2} \bigg( 6\, (d_e^A)^3 - \sqrt{2}\, G_F\, g_{\mu \tau} \big(C_A^{\mu,\text{SM}} - C_A^{\tau,\text{SM}} \big)\, q_r^2 + G_F\, d_e^A \Big[ g_{\mu \tau}^2 - 2 \sqrt{2}\, G_F\,q^2_r \, \sum_\alpha C_A^{\alpha,\text{SM}}  \Big]  \bigg)\,.
\end{split}
\end{equation}
}
\normalsize
\end{widetext}

\begin{figure}[b]
    \begin{center}
    \includegraphics[width=0.5\textwidth]{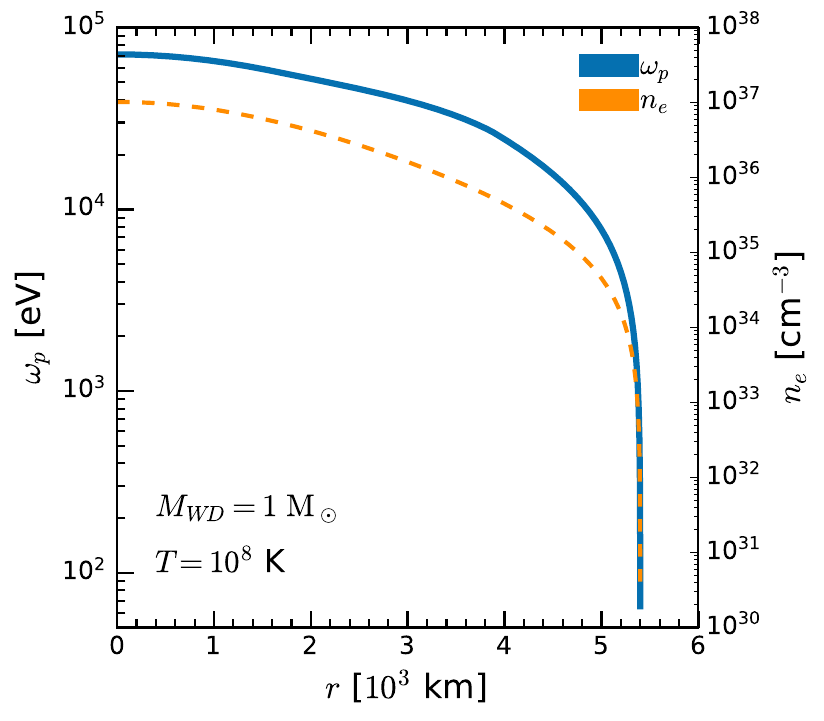}
    \end{center}
    \caption{Plasma frequency $\omega_p$ (blue line) and electron density $n_e$ (orange dashed line) of a WD  with \mbox{$M_\mathrm{WD}=1 M_\odot$} at {$T_\mathrm{WD}= 10^8$ K} as a function of the distance to the centre of the star in km.}
    \label{fig:plamsa_freq}
\end{figure}

\subsection{Resonant dark photons} 
\label{sec:resonant-region}

When the dark photon mass is roughly of the same order as the WD temperature of~$\mathcal{O}$(keV) the $A'$ contribution to the WD emissivities  are significantly enhanced due to a resonance in the $A'$ propagator. More precisely, this happens when the dark photon mass hits the plasma frequency, which is defined as~\cite{Braaten:1993jw}
{\small
\begin{equation}
\label{eq:plasma_freq}
\begin{split}
\omega_p^2 = &\frac{4\alpha}{\pi} \int_0^\infty d|\tv{k}| \frac{\tv{k}^2}{E_k} \bigg(1 - \frac{1}{3} v^2 \bigg) \big( f_e (E_k) + f_{\overline{e}} (E_k) \big)\,,
\end{split}
\end{equation}
}
\normalsize
where $f_e$ and $f_{\overline{e}}$ are the Fermi thermal distributions for electrons and positrons, respectively. We are integrating over the $3-$momentum of the electrons and positrons in the plasma with energy {$E_k \equiv \sqrt{\tv{k}^2 + m_e^2}$} and velocity $v \equiv \tv{k} / E_k$. The plasma frequency $\omega_p$ depends on the temperature \Twd and the chemical potential $\mu$, which in turn depends on the distance $r$ from the centre of the WD. 

In \cref{fig:plamsa_freq}, we show  the plasma frequency $\omega_p$ and electron density $n_e$ as a function of the WD radius $r$. We can see that the plasma frequency reaches its maximum at the centre of the star and decreases roughly over an order of magnitude throughout the interior of the star before it rapidly drops in the outer layers.
This behaviour can be readily understood by looking at the electron density profile since the plasma frequency $\omega_p$ is directly proportional to it. 

Thus, the resonance region consists of a whole range of dark photon masses for which a pole arises in the $A^\prime$ propagator due to the scanning of the plasma frequency $\omega_p$ in the range \mbox{$0 \le r \le R_\mathrm{WD}$}.
This effect, if not cured, leads to a continuous curve of divergences of the integrands in~\cref{eq:QL,eq:QT,eq:QA} in the $|\tv{q}|-r$ plane, where $\tv{q}$ is the $3-$momentum of the external plasmon and $r$ the distance from the centre of the WD.

However, these divergences are non-physical and can be cured by considering the Breit-Wigner (BW) propagator~\cite{Breit:1936zzb}. This takes into account the imaginary component of the self-energy, which is directly related to the instability of the particle, and therefore is closely related to its decay width.  The BW propagator takes the form~\cite{Weldon:1983jn}
\begin{equation}\label{eq:bw_prop}
\begin{split}
G_{\mathrm{BW}}^{\mu \nu} (q^2) = &\frac{-i (g^{\mu \lambda} - q^\mu q^\lambda / m^2)}{q^2 - m^2 - \mathrm{Re}(F) - i\, \mathrm{Im}(F)} P_{L \lambda}^{\nu} \\& \ + \frac{-i (g^{\mu \lambda} - q^\mu q^\lambda / m^2)}{q^2 - m^2 - \mathrm{Re}(G) - i\, \mathrm{Im}(G)} P_{T \lambda}^{\nu} \, ,
\end{split}
\end{equation}
where the self-energy of the vector is \mbox{$\Pi^{\mu \nu} = F\, P_L^{\mu \nu} + G\, P_T^{\mu \nu}$}, in terms of the projectors $P_L^{\mu \nu}$ and $P_T^{\mu \nu}$ given in ~\cref{eq:projs}.

There are, in principle, two main contributions to the imaginary part of the dark photon self energy $\mathrm{Im}(\Pi_{A'})$. The first one is due to a thermal loop of electrons, while the second one is due to a zero-temperature loop of neutrinos, since neutrinos are not thermalized inside the WD star. In the resonance region of masses $m_{A'} \ll1$ MeV, there is no imaginary contribution from the electron loop since the dark photon is not massive enough to decay into an electron-positron pair. It can, however, decay into a pair of neutrino-antineutrino, which for the corresponding $A'$ masses and WD temperatures can be considered to be massless. We have to include the imaginary contributions $i\, \mathrm{Im}(\Pi_{A^\prime,\lambda})$ in the propagators of each polarisation $\lambda = T, L$ in \cref{eq:full_prop_dark_photon}.

After renormalizing the self-energy with internal neutrinos of flavour $\alpha$ using the $\overline{\mathrm{MS}}$ scheme, we obtain the following expressions,
\begin{equation}
\begin{split}
\bar{\Pi}^{\mu \nu}_{A^\prime} (q^2) = - &\frac{\big(k_\nu^\alpha\big)^2}{4\pi^2} q^2 g^{\mu \nu} \int_0^1 d{x} \, x \, (1-x) \\&\times \log \bigg(\frac{m_\alpha^2}{m_\alpha^2 - x(1-x)q^2} \bigg) \,,
\end{split}
\end{equation}
where $k_\nu^\alpha$ is the coefficient defined in \cref{eq:dark_coeffs}.
In the limit of massless neutrinos, the imaginary part of this self-energy is easily found as the argument of the logarithm is always negative in the region of integration, since in turn the momentum of the thermal on-shell plasmon, $q^2$, is always positive. Hence, we find
\begin{equation}
\begin{split}
\mathrm{Im}(\bar{\Pi}^{\mu \nu}_{A^\prime}) (q^2) &= \frac{\big(k_\nu^\alpha\big)^2}{24\pi} q^2 g^{\mu \nu}\\
&= \frac{\big(k_\nu^\alpha\big)^2}{24\pi} \frac{(\omega_l^2 - \tv{q}^2)^2}{\tv{q}^2} P_L^{\mu \nu} \\& \ \ - \frac{\big(k_\nu^\alpha\big)^2}{24\pi} (\omega_t^2 - \tv{q}^2) P_T^{\mu \nu} \,,
\end{split}
\end{equation}
where we have replaced $q^2$ by its explicit value for the two polarizations.

Finally, the quantities entering the denominators of the dark photon propagator in~\cref{eq:bw_prop}, following the same procedure as in \cref{sec:ultra-light-A}, are \mbox{$\mathrm{Im}(F)=\mathrm{Im}(\bar{\Pi}^{0 0}_{A^\prime}) \times q^2 / \tv{q}^2$} and \mbox{$\mathrm{Im}(G)=\mathrm{Im}(\bar{\Pi}^{x x}_{A^\prime})$ }for the longitudinal and transverse polarisation, respectively. The projectors are absorbed when multiplied by those of the electronic loop of the plasmon. To obtain the full expression we still have to sum over the different neutrino flavours $\alpha$ in $\mathrm{Im}(\bar{\Pi}^{\mu \nu}_{A^\prime})$,  \mbox{$k_\nu^\alpha \to \sum_\alpha k_\nu^\alpha$}.

This prescription regulates the divergences in the dark photon propagator and leads to finite expressions in the whole parameter space. Nevertheless, the curve of very narrow peaks in the $|\tv{q}| - r$ plane of integration remains. To reliably evaluate the luminosity integrals in this regime, we have to thoroughly sample these narrow peaks, for which we were relying on the \texttt{VEGAS+} algorithm~\cite{Lepage:2020tgj} for adaptive multidimensional Monte Carlo integration in our calculations.

\section{Results}
\label{sec:limits}

\begin{figure*}[!th]
    \begin{center}
    \includegraphics[width=1.\textwidth]{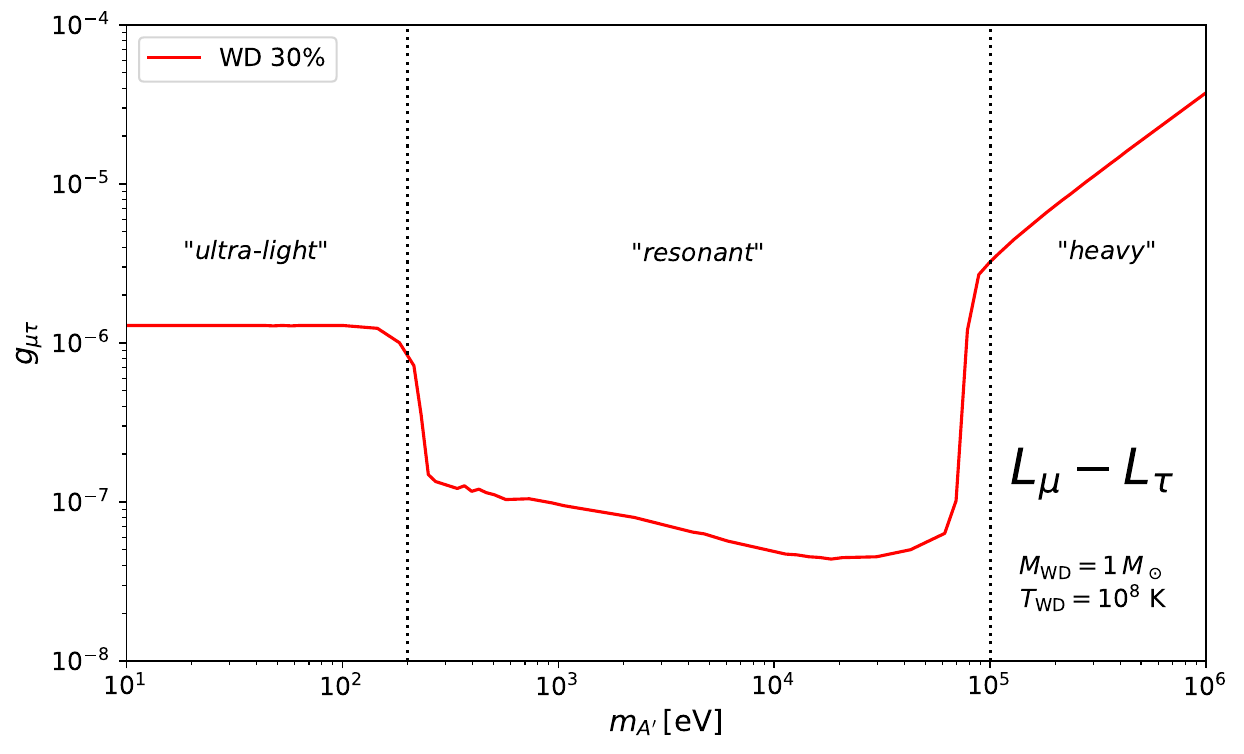}
    \end{center}
    \caption{WD cooling bounds on sub-MeV  \Umt bosons. At high dark photon masses of $m_{A'} \gtrsim 100$ keV, the $A'$ propagator in the WD emissivities is dominated by the $A'$ mass and is to good approximation independent of the plasma frequency $\omega_p$. Below this threshold the $A'$ mass scans the plasma frequency in different phase space regions, leading to a resonance behaviour in the window of $0.2$ keV  $\lesssim m_{A'} \lesssim 100$ keV. At the low-mass end of this window the resonance peaks of the BW propagator start slowly moving outside the integration domain of the luminosity $L_\mathrm{WD}$ towards higher radii $r>R_\mathrm{WD}$ resulting in an attenuation of the resonance. Below $m_{A'}\lesssim 200$ eV the integrand no longer exposes any resonance behaviour in the integration domain and the integrand is dominated by the plasma frequency and approximately independent of the $A'$ mass.}
    \label{fig:lims_full}
\end{figure*}

In the following, we present our results for WD cooling via plasmon decay in presence of a \Umt gauge boson. The results were derived for a representative WD with a  mass  of $M_\mathrm{WD} = 1 \ M_\odot$, a temperature of \Twd$=10^8$ K, and the profiles of the electron number density $n_e$ and chemical potential $\mu$ obtained from solving the Salpeter EoS~\cite{Salpeter:1961zz}. For the numerical computation of the self-energy, we have considered two main approximations. For the inner layers of the WD, we considered a degenerate regime, where the chemical potential $\mu$ dominates over the temperature \Twd and the electron mass $m_e$. The quantities depend mainly on the Fermi momentum in this regime. For the outer layers, we regarded the classical regime, where the mass of the electron dominates over the temperature and over the sum of the chemical potential and temperature. In this regime, the quantities depend mainly on the electron number density $n_e$. More details on these limits and their respective expressions can also be found in \cite{Salpeter:1961zz}.

In~\cref{fig:lims_full} we present the general behaviour of the WD cooling limits due to plasmon decay via an extra \Umt boson  obtained by the computations outlined in~\cref{sec:wd_cooling}. 
Current best fits of the hot WD neutrino luminosity function allow for a variation in the neutrino cooling with respect to the SM by a relative factor of $0.66\lesssim f_s \lesssim 1.31$ at the $90\, \%$ confidence level (CL)  (see~\cref{app:limit} for the extraction of this limit based on Ref.~\cite{Hansen:2015lqa}).
Therefore, we can exclude parameter space leading to more than $30\,\%$ (red line) WD excess cooling in neutrinos.
We show the limit of $\varepsilon^{\text{BSM}} = 0.3$ extra cooling relative to the SM in the $m_{A^\prime} - g_{\mu \tau}$ plane. This plot illustrates well the three qualitatively different domains of WD cooling depending on the $A'$ mass already outlined throughout the paper. 

For heavy dark photon masses of \mbox{$m_{A'} \gtrsim 10^5$ eV}, we observe a linear scaling of the sensitivity line. This is readily explained by looking at the WD plasma frequency in~\cref{fig:plamsa_freq}. Throughout the star the plasma frequency assumes its maximum value at \mbox{$\omega_{p,\mathrm{max}} \lesssim 10^5$ eV}. Hence, for dark photon masses above $\omega_{p,\mathrm{max}}$ the $A'$ propagator in the emissivity is entirely dominated by its mass and scales as $\sim 1/ m_{A^\prime}^2$.  The linear increase can be understood from the solution of the approximation in \cref{eq:epsilon_heavy}.

On the other hand, for ultra-light dark photons instead, we see a constant scaling of the sensitivity independent of the dark photon mass. This can be understood by an inspection of the propagator of the dark photon. As we stated in section \cref{sec:ultra-light-A}, for ultra-light dark photons with $m_{A'}\ll q$ much smaller than the typical momenta in the stellar interior, we can well approximate the propagator in the WD emissivities by $\sim 1/ q^2$. Hence, in this very light regime the WD sensitivity becomes independent of the $A'$ mass to excellent approximation. 

Finally, we will discuss the central region of the plot where the resonance behaviour dominates. For masses of the order of the plasma frequency, $\omega_p$, there is a resonance in the propagator between the dark photon mass $m_{A'}$ and the momentum transfer $q$. As can be seen in \cref{fig:plamsa_freq}, as soon as the dark photon masses fall below the maximum plasma frequency of $\omega_{p,\mathrm{max}}\lesssim 10^5$ eV, resonance will occur in some region in the interior of the star. This explains why the transition between the linear and resonant regime is so abrupt. Employing the Breit-Wigner prescription, this will result in narrow but regulated peaks in the emissivities. Since these peaks contribute significantly to the integral, they are largely enhancing the sensitivity of WD cooling to a new \Umt boson. As the dark photon masses decrease, the location of these peaks within the integration domain will shift towards higher and higher radii. For masses below about $\sim 300$ eV these peaks will start moving past the WD radius $R_\mathrm{WD}$ and out of the integration domain. Hence, their contribution becomes less and less important until no resonance occurs at all for  masses of {$m_{ A^\prime } \lesssim 200$ eV} and the ultra-light regime is reached.

\begin{figure*}
    \centering
    \includegraphics[width=1.\textwidth]{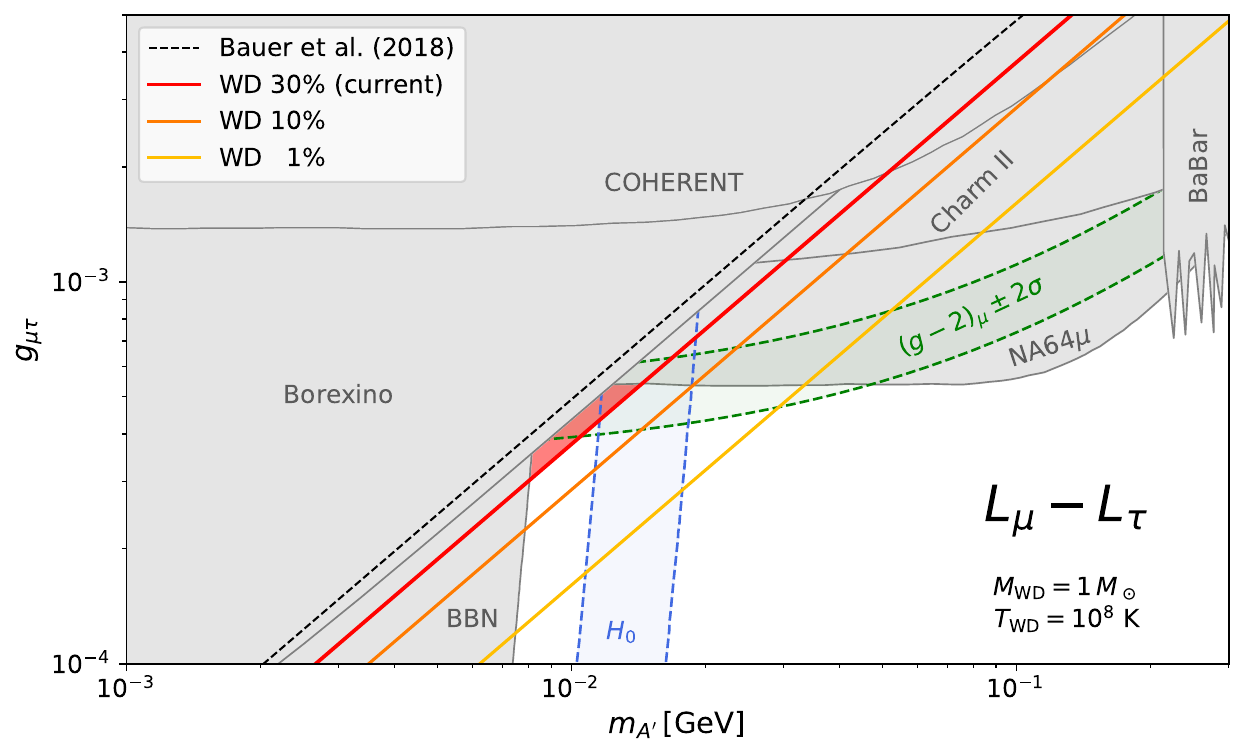}
    \caption{Limits on MeV-mass \Umt gauge bosons. The grey shaded areas show the current constraints (see text for explanations). The green and blue bands depict the regions of parameter space preferred by the \gmu and $H_0$ anomaly, respectively. We show our constraints from white dwarf cooling via plasmon decay for an excess of  30\% (red), 10\% (orange) and 1\% (yellow) over the SM neutrino cooling rate. The red coloured area shows newly excluded parameter space from the 30\% excess cooling limit of this analysis. For comparison, the black dashed line shows the previously estimated WD cooling bound of Ref.~\cite{Bauer:2018onh}.}
    \label{fig:lims}
\end{figure*}

In~\cref{fig:lims} we show the WD cooling limit on a \Umt boson in the MeV mass window, where a solution of the muon \gmu anomaly is still allowed. 
For comparison, we show the current best limits in grey. At masses below \mbox{$\mathcal{O}(10)$ MeV} the dark photon $A'$ contributes significantly to the heating of the neutrino gas in the early universe resulting in a too large number of neutrino degrees of freedom, $\Delta N_\mathrm{eff}$, during BBN~\cite{Escudero:2019gzq}.
However, in the mass window of  $m_{A'} \sim 10 - 20$ MeV
the same effect leads to a milder contribution to $N_\mathrm{eff}$, which could explain the Hubble tension~\cite{Escudero:2019gzq}. The corresponding favoured region is depicted by the blue band labelled $H_0$. 
In turn, the green band shows the region of parameter space preferred by the \gmu anomaly as measured at the E989 experiment~\citep{Muong-2:2023cdq}  and
explained in~\cref{sec:g2}. The most stringent existing constraint on the \gmu favoured region is due to the recently reported result of the invisible search at the NA64$\mu$ experiment~\cite{Andreev:2024sgn}.
At low masses constraints from the measurement of the {$^7$Be} solar neutrino flux at the Borexino experiment~\cite{Bellini:2011rx,Borexino:2017rsf,Amaral:2020tga} exclude a solution below $A'$ masses of $m_{A'}\sim 10$ MeV.
At high masses, we also show the limit obtained from resonance searches in four-muon production by the BaBar collaboration~\cite{BaBar:2016sci}.
For comparison, we also show the resulting bound from measurements of coherent elastic neutrino-nucleus scattering (\cevns) with a CsI[Na] target at the COHERENT experiment~\cite{COHERENT:2015mry,COHERENT:2017ipa}.
Similarly, a strong constraint arises from the search for neutrino trident production~\cite{Altmannshofer:2014pba}. We show the leading bound by the CHARM-II experiment~\cite{CHARM-II:1990dvf}.\footnote{In principle, a more stringent bound can be derived from the CCFR results~\cite{Altmannshofer:2014pba}. However, some doubts have been shed on the correct incorporation of a background due to diffractive charm production in the relevant analysis~\cite{Krnjaic:2019rsv} and hence we abstain from showing the corresponding limit.}

The solid red line is the novel constraint derived in this work from excluding $\varepsilon^{\text{BSM}}=30\,\%$ extra WD neutrino cooling. 
For comparison, the black dashed line shows the estimate of the WD cooling bound that has been obtained in Ref.~\cite{Bauer:2018onh} from matching to the EFT coefficients used in the analysis of~\cite{Dreiner:2013tja}, which in turn assumed a 100\% excess cooling constraint.
We note that the current 30\% WD cooling limit excludes part of the parameter space favoured by a simultaneous explanation of \gmu and $H_0$, which  previously has not been tested (red area).

We also show the sensitivities of a future measurement of the hot WD neutrino luminosity function excluding extra cooling at the 10\% (orange line) and 1\% (yellow line) level. 
The future space-based CASTOR experiment~\cite{10.1117/12.926198} is planning to observe approximately 8 million hot young WDs~\cite{Fantin_2020}. Hence, CASTOR is expected to collect a factor of $\sim 10^3$ more statistics as compared to the sample used in Ref.~\cite{Hansen:2015lqa}, which was used to extract the current 30\% excess cooling bound. CASTOR is expected to be launched as early as 2027 with a minimum lifetime of the spacecraft of 5 years~\cite{castor19}, so that an order of magnitude improvement in the WD excess cooling bound within the next decade seems feasible. Improved sensitivities at the level of 10\% (orange line) excess cooling could rule out almost all of the joint \gmu and $H_0$ explanation, while sensitivities at the level of $1\%$ (yellow line) could exclude almost the entire parameter space favoured by the \gmu anomaly.

\section{Conclusions}
\label{sec:conclusions}

In this article, we have performed for the first time an \textit{ab initio} computation of the WD luminosity due to plasmon decay into neutrinos in presence of an extra new \Umt gauge boson. We have performed the calculations for a representative WD star with a mass of $M_\mathrm{WD} = 1 M_\odot$ and temperature of $\Twd = 10^8$ K. Following \cite{Braaten:1993jw}, we have considered two main approximations: the degenerate regime (for the inner layers of the WD) and the classical regime (for the outer ones). 

Our main result is a careful derivation of the neutrino luminosity for the entire mass range of a \Umt boson resulting in the limits shown in~\cref{fig:lims_full}. In particular, we have also performed the computation in the
resonant regime where the $A'$ masses are comparable to the plasma frequency $\omega_p$. We have demonstrated that the varying profile of the electron density $n_e$ and chemical potential $\mu$ in the interior of the star result in a broad range of $A'$ masses of \mbox{100 eV $\lesssim m_{A'} \lesssim 100$ keV}, in which plasma resonance effects become important and lead to a significantly increased cooling contribution of the $A'$.

We have further demonstrated that the current bound of 30\% extra WD cooling is the leading constraint on the region of parameter space favoured by a simultaneous explanation of the \gmu and $H_0$ anomaly (cf. red line in~\cref{fig:lims}). A future precision observation of the WD neutrino luminosity function with CASTOR at the 10\% or even 1\% level could exclude a large fraction or even the entire remaining parameter space favoured by \gmu.

A straightforward extension of this work would be to perform the same calculations for neutron stars. However, the lack of knowledge of the precise equation of state for these stars makes it fundamentally more difficult to obtain robust results for the corresponding luminosities. Nevertheless, compared to WDs these objects exhibit much higher densities and at their birth also much higher temperatures of up to $T\sim 50$ MeV. This could significantly enhance the effect of neutrino cooling through a novel \Umt boson and in particular improve sensitivities at higher boson masses. Furthermore, resonances at a higher mass range could result in limits improving over current ones for mediators like leptophilic dark photons.

\section*{Acknowledgements}

We would like to thank Maura E. Ramirez-Quezada for the profiles of the white dwarf used for the computation of the luminosities of this paper. We also want to acknowledge the \texttt{VEGAS+} package~\cite{peter_lepage_2024_10783443} for multidimensional Monte Carlo integration.
This research project was made possible through the access granted by the Galician Supercomputing Center (CESGA) to its supercomputing infrastructure. The supercomputer FinisTerrae III and its permanent data storage system have been funded by the Spanish Ministry of Science and Innovation, the Galician Government and the European Regional Development Fund (ERDF).

The work of PF was supported by  the Spanish Agencia Estatal de Investigaci\'on through the grants PID2021-125331NB-I00 and CEX2020-001007-S, funded by
MCIN/AEI/10.13039/501100011033.
The research of JHZ has received support from the European Union’s Horizon 2020 research and innovation programme under the Marie Sk\l{}odowska-Curie grant  agreement No 860881-HIDDeN.

\appendix

\section{Model details} 
\label{app:model}

\subsection{Loop-induced kinetic mixing in \Uij}
\label{app:mixing}

In this section, we discuss the computation of the one-loop induced kinetic mixing. In this work we are mainly interested in \Umt, however, this discussion is generic for all \Uij with $i,j=e,\mu,\tau$.

Since part of the SM leptons are charged under both the SM hypercharge $U(1)_Y$ and the new leptophilic gauge group \Uij, a kinetic mixing term between the new gauge boson and the hypercharge boson as well as the neutral $SU(2)_L$ boson is induced at the one-loop level~\cite{Bauer:2022nwt},
\begin{widetext}
\begin{align}\label{eq:liljmix_B}
\epsilon_B^{ij}(q^2) &=\frac{g'\, g_{ij}}{8\pi^2}\int_0^1 dx\,x(1-x)\, \left[ 3 \log\left(\frac{m_i^2-x(1-x)q^2}{m_j^2-x(1-x)q^2}\right)  + \log\left(\frac{m_{\nu_i}^2-x(1-x)q^2}{m_{\nu_j}^2-x(1-x)q^2}\right)  \right] \,, \\
\epsilon_W^{ij}(q^2) &=\frac{g\; g_{ij}}{8\pi^2}\int_0^1 dx\,x(1-x)\, \left[ \log\left(\frac{m_i^2-x(1-x)q^2}{m_j^2-x(1-x)q^2}\right)  - \log\left(\frac{m_{\nu_i}^2-x(1-x)q^2}{m_{\nu_j}^2-x(1-x)q^2}\right)  \right] \,.
\label{eq:liljmix_W}
\end{align}
\end{widetext}
In the physical basis of neutral photon and $Z$ boson, the mixings with the new leptophilic gauge boson read
\begin{widetext}
\begin{align}\label{eq:liljmix_A}
\epsilon_A^{ij}(q^2) &=\frac{e\, g_{ij}}{2\pi^2}\int_0^1 dx\,x(1-x)\ \log\left(\frac{m_i^2-x(1-x)q^2}{m_j^2-x(1-x)q^2}\right)  \,, \\
\epsilon_Z^{ij}(q^2) &= -\frac{g_z\, g_{ij}}{4\pi^2}\int_0^1 dx\,x(1-x)\, \left[ \log\left(\frac{m_i^2-x(1-x)q^2}{m_j^2-x(1-x)q^2}\right)  +  \log\left(\frac{m_{\nu_i}^2-x(1-x)q^2}{m_{\nu_j}^2-x(1-x)q^2}\right)  \right] \,.
\label{eq:liljmix_Z}
\end{align}
\end{widetext}

Since the typical core temperature of a white dwarf is of the order of $\sim$ 10 keV, it is safe to assume that the typical energy transfer $q^2$ at which the plasmon decay is happening, is much larger than the neutrino masses, $q^2\gg m_\nu^2$. Thus, we can expand the second term in~\cref{eq:liljmix_Z},
\begin{align}
\log\left(\frac{{m_{\nu_i}^2}/{q^2}-x(1-x)}{m_{\nu_j}^2/q^2-x(1-x)}\right) \approx \log\left(\frac{x(1-x)}{x(1-x)}\right) =0   \,,
\end{align}
where the first approximation is valid almost everywhere in the interval $x\in[0,1]$.

At the same time, the masses of the charged leptons are much larger than the WD temperatures, $q^2\ll m_\ell^2$. Hence, we can also expand the logarithms containing the charged lepton masses in~\cref{eq:liljmix_A,eq:liljmix_Z}, 
\begin{align}
\log\left(\frac{m_i^2/m_j^2-x(1-x)\ q^2/m_j^2}{1-x(1-x)\ q^2/m_j^2}\right) \approx \log\left(\frac{m_i^2}{m_j^2}\right)  \,,
\end{align}

In summary, at typical WD temperatures, we can work with the $q^2$-independent approximations,
\begin{align}\label{eq:liljmix_A_approx}
\epsilon_A^{ij} &\approx\frac{e\, g_{ij}}{6\pi^2}\ \log\left(\frac{m_i}{m_j}\right)  \,, \\
\epsilon_Z^{ij} &\approx -\frac{g_z\, g_{ij}}{12\pi^2}\ \log\left(\frac{m_i}{m_j}\right) = - \frac{1}{2}\frac{s_W}{c_W} \,\epsilon_A^{ij}  \,.
\label{eq:liljmix_Z_approx}
\end{align}

\subsection{Interaction terms of the boson mass eigenstates}
\label{app:interactions}

In order to obtain the interaction terms of the physical mass eigenstates of the neutral gauge bosons, we have to diagonalise both the kinetic and the mass terms in~\cref{eq:full_lag}.  Applying the full diagonalisation to the interaction terms, we obtain
\begin{widetext}
\begin{align} \label{eq:interaction_full}
\mathcal{L}_\mathrm{int}  = - (A_{\mu}, Z_\mu, A'_{\mu})\,
    \begin{pmatrix}
    e \, j_\mathrm{EM}^\mu \\[3ex]
    \frac{1}{\tau} \sqrt{\frac{\kappa-1}{2\, \kappa}} \left[g_x \, j^\mu_x
    -e\, \epsilon_A \, j_\mathrm{EM}^\mu -g_z \left(\epsilon_Z-\tau \sqrt{\frac{\kappa+1}{\kappa-1}}\right) j^\mu_Z \right]\\[3ex]
    \frac{1}{\tau} \sqrt{\frac{\kappa+1}{2\, \kappa}} \left[g_x \, j^\mu_x
    -e\, \epsilon_A \, j_\mathrm{EM}^\mu -g_z \left(\epsilon_Z+\tau \sqrt{\frac{\kappa-1}{\kappa+1}}\right) j^\mu_Z \right]
    \end{pmatrix}  \,, 
\end{align}
\end{widetext}
with the definitions
\begin{align}
\tau & =\sqrt{1-\epsilon_A^2-\epsilon_Z^2}\,, \\
\kappa & =  \sqrt{1+4 \left(\frac{\epsilon_Z}{\delta-1}\right)^2}\,, \\
\delta & = \left(\frac{M_X}{M_Z}\right)^2\,.
\end{align}

Since we are considering the mass range $M_X\ll M_Z$ we can approximate $\delta\approx 0$. Furthermore, we are mostly interested in the parameter space where $g_x<10^{-2}$, such that $\epsilon_A\lesssim10^{-4}$. To linear order in the the small parameters $\epsilon_A$ and $\epsilon_Z$, we then get 
{\small
\begin{align} \label{eq:interaction_approx}
\mathcal{L}_\mathrm{int}  \approx - (A_{\mu}, Z_\mu, A'_{\mu})\,
    \begin{pmatrix}
    e \, j_\mathrm{EM}^\mu \\[4pt]
    g_z\, j^\mu_Z\\[4pt]
    g_x \, j^\mu_x
    -e\, \epsilon_A \, j_\mathrm{EM}^\mu -g_z \,\epsilon_Z\, j^\mu_Z 
    \end{pmatrix}  \,.
\end{align}
}
\normalsize
Hence, we recover the familiar SM photon and $Z$ interactions with the electromagnetic and $Z$ currents as defined in~\cref{eq:emcurr,eq:zcurr}, respectively. The mass eigenstate of the \Umt gauge boson $A'$, however, not only couples to the \Umt current but picks up an additional interaction with the electromagnetic and $Z$ current suppressed by the kinetic mixings $\epsilon_A$ and $\epsilon_Z$, respectively.

\section{Limit extraction} 
\label{app:limit}

\begin{figure}[b]
    \begin{center}
    \includegraphics[width=0.5\textwidth]{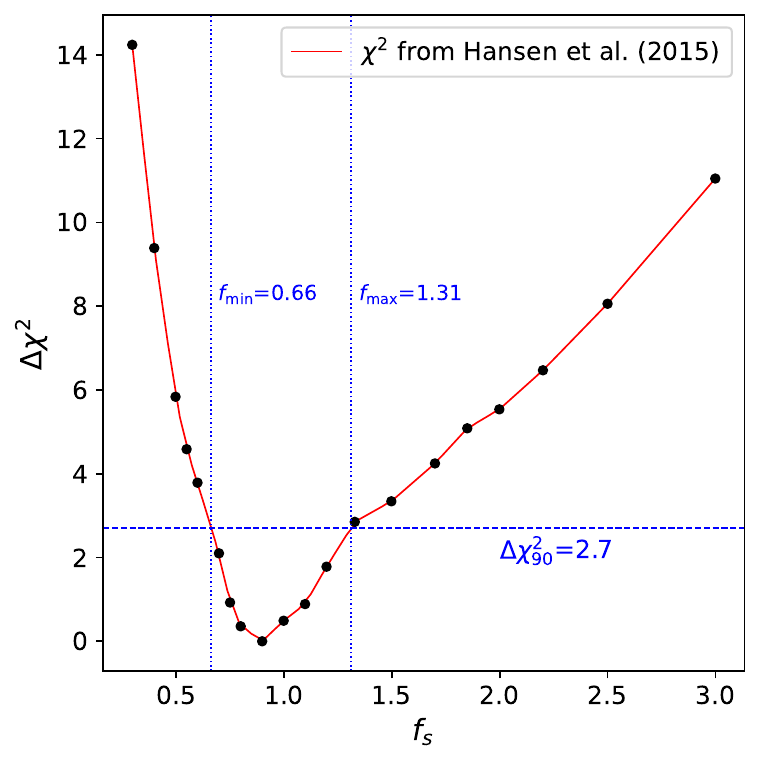}
    \end{center}
    \caption{We use the $\chi^2$ function computed in Ref.~\cite{Hansen:2015lqa} to extract the limit at the 90\% CL ($\Delta\chi^2=2.7$) for modified cooling by a factor of $f_s$ relative to the SM.}
    \label{fig:chisq}
\end{figure}

To set a limit on extra WD cooling through neutrinos, we utilise the $\chi^2$ function provided in Fig.~10 of Ref.~\cite{Hansen:2015lqa}. The plot in~\cref{fig:chisq} shows $\Delta\chi^2$ as a function of the linear scaling parameter $f_s$ quantifying extra neutrino cooling relative to the SM. The original limit provided in~\cite{Hansen:2015lqa} was set at the 95\% CL and resulted in an allowed interval of $0.6\lesssim f^{95}_s \lesssim 1.7$. However, in order to be consistent with the literature on dark photons, where limits in the coupling-versus-mass plane are typically set at the 90\% CL, we also extract the 90\% CL limit on the scaling parameter $f_s$ by setting $\Delta\chi^2=2.7$ (for 1 dof) in the $\chi^2$ function of~\cite{Hansen:2015lqa}.  As illustrated by the two blue vertical lines in~\cref{fig:chisq}, this results in an interval of $0.66\lesssim f^{90}_s \lesssim 1.31$.
Hence, we can exclude 30\% extra neutrino cooling at the 90\% CL (cf.~red line in~\cref{fig:lims_full,fig:lims}) given the analysis of~\cite{Hansen:2015lqa}.

\bibliographystyle{utphys}
\bibliography{literature}

\providecommand{\href}[2]{#2}\begingroup\raggedright\begin{thebibliography}{10}

\bibitem{Foot:1990mn}
R.~Foot, ``{New Physics From Electric Charge Quantization?},''
  \href{http://dx.doi.org/10.1142/S0217732391000543}{{\em Mod. Phys. Lett. A}
  {\bf 6} (1991)  527--530}.

\bibitem{He:1990pn}
X.~G. He, G.~C. Joshi, H.~Lew, and R.~R. Volkas, ``{NEW Z-prime
  PHENOMENOLOGY},'' \href{http://dx.doi.org/10.1103/PhysRevD.43.R22}{{\em Phys.
  Rev. D} {\bf 43} (1991)  22--24}.

\bibitem{He:1991qd}
X.-G. He, G.~C. Joshi, H.~Lew, and R.~R. Volkas, ``{Simplest Z-prime model},''
  \href{http://dx.doi.org/10.1103/PhysRevD.44.2118}{{\em Phys. Rev. D} {\bf 44}
  (1991)  2118--2132}.

\bibitem{Foot:1994vd}
R.~Foot, X.~G. He, H.~Lew, and R.~R. Volkas, ``{Model for a light Z-prime
  boson},'' \href{http://dx.doi.org/10.1103/PhysRevD.50.4571}{{\em Phys. Rev.
  D} {\bf 50} (1994)  4571--4580},
  \href{http://arxiv.org/abs/hep-ph/9401250}{{\tt arXiv:hep-ph/9401250}}.

\bibitem{Heeck:2011wj}
J.~Heeck and W.~Rodejohann, ``{Gauged $L_\mu - L_\tau$ Symmetry at the
  Electroweak Scale},''
  \href{http://dx.doi.org/10.1103/PhysRevD.84.075007}{{\em Phys. Rev. D} {\bf
  84} (2011)  075007}, \href{http://arxiv.org/abs/1107.5238}{{\tt
  arXiv:1107.5238 [hep-ph]}}.

\bibitem{Asai:2018ocx}
K.~Asai, K.~Hamaguchi, N.~Nagata, S.-Y. Tseng, and K.~Tsumura, ``{Minimal
  Gauged U(1)$_{L_\alpha - L_\beta}$ Models Driven into a Corner},''
  \href{http://dx.doi.org/10.1103/PhysRevD.99.055029}{{\em Phys. Rev. D} {\bf
  99} (2019) no.~5, 055029}, \href{http://arxiv.org/abs/1811.07571}{{\tt
  arXiv:1811.07571 [hep-ph]}}.

\bibitem{Bauer:2020itv}
M.~Bauer, P.~Foldenauer, and M.~Mosny, ``{Flavor structure of anomaly-free
  hidden photon models},''
  \href{http://dx.doi.org/10.1103/PhysRevD.103.075024}{{\em Phys. Rev. D} {\bf
  103} (2021) no.~7, 075024}, \href{http://arxiv.org/abs/2011.12973}{{\tt
  arXiv:2011.12973 [hep-ph]}}.

\bibitem{Majumdar:2020xws}
C.~Majumdar, S.~Patra, P.~Pritimita, S.~Senapati, and U.~A. Yajnik, ``{Neutrino
  mass, mixing and muon g \ensuremath{-} 2 explanation in $
  \mathrm{U}{(1)}_{L_{\mu }-{L}_{\tau }} $ extension of left-right theory},''
  \href{http://dx.doi.org/10.1007/JHEP09(2020)010}{{\em JHEP} {\bf 09} (2020)
  010}, \href{http://arxiv.org/abs/2004.14259}{{\tt arXiv:2004.14259
  [hep-ph]}}.

\bibitem{Singh:2022tvz}
L.~Singh, M.~Kashav, and S.~Verma, ``{Gauged
  U(1)$L_\ensuremath{\mu}\ensuremath{-}L_\ensuremath{\tau}$ symmetry and
  two-zero textures of inverse neutrino mass matrix in light of muon (g
  \ensuremath{-} 2)},'' \href{http://dx.doi.org/10.1142/S0217732322502029}{{\em
  Mod. Phys. Lett. A} {\bf 37} (2022) no.~30, 2250202},
  \href{http://arxiv.org/abs/2207.08415}{{\tt arXiv:2207.08415 [hep-ph]}}.

\bibitem{Arora:2022hza}
S.~Arora, M.~Kashav, S.~Verma, and B.~C. Chauhan, ``{Muon (g \ensuremath{-} 2)
  in U(1)$_{L_\mu - L_\tau}$ scotogenic model extended with vector like
  fermion},'' \href{http://dx.doi.org/10.1088/1402-4896/acb32b}{{\em Phys.
  Scripta} {\bf 98} (2023) no.~2, 025304},
  \href{http://arxiv.org/abs/2206.12828}{{\tt arXiv:2206.12828 [hep-ph]}}.

\bibitem{Baek:2008nz}
S.~Baek and P.~Ko, ``{Phenomenology of U(1)(L(mu)-L(tau)) charged dark matter
  at PAMELA and colliders},''
  \href{http://dx.doi.org/10.1088/1475-7516/2009/10/011}{{\em JCAP} {\bf 10}
  (2009)  011}, \href{http://arxiv.org/abs/0811.1646}{{\tt arXiv:0811.1646
  [hep-ph]}}.

\bibitem{Baek:2015fea}
S.~Baek, ``{Dark matter and muon $(g-2)$ in local
  $U(1)_{L_\mu-L_\tau}$-extended Ma Model},''
  \href{http://dx.doi.org/10.1016/j.physletb.2016.02.062}{{\em Phys. Lett. B}
  {\bf 756} (2016)  1--5}, \href{http://arxiv.org/abs/1510.02168}{{\tt
  arXiv:1510.02168 [hep-ph]}}.

\bibitem{Biswas:2016yan}
A.~Biswas, S.~Choubey, and S.~Khan, ``{Neutrino Mass, Dark Matter and Anomalous
  Magnetic Moment of Muon in a $U(1)_{L_{\mu}-L_{\tau}}$ Model},''
  \href{http://dx.doi.org/10.1007/JHEP09(2016)147}{{\em JHEP} {\bf 09} (2016)
  147}, \href{http://arxiv.org/abs/1608.04194}{{\tt arXiv:1608.04194
  [hep-ph]}}.

\bibitem{Biswas:2016yjr}
A.~Biswas, S.~Choubey, and S.~Khan, ``{FIMP and Muon ($g-2$) in a
  U$(1)_{L_{\mu}-L_{\tau}}$ Model},''
  \href{http://dx.doi.org/10.1007/JHEP02(2017)123}{{\em JHEP} {\bf 02} (2017)
  123}, \href{http://arxiv.org/abs/1612.03067}{{\tt arXiv:1612.03067
  [hep-ph]}}.

\bibitem{Foldenauer:2018zrz}
P.~Foldenauer, ``{Light dark matter in a gauged $U(1)_{L_\mu-L_\tau}$ model},''
  \href{http://dx.doi.org/10.1103/PhysRevD.99.035007}{{\em Phys. Rev. D} {\bf
  99} (2019) no.~3, 035007}, \href{http://arxiv.org/abs/1808.03647}{{\tt
  arXiv:1808.03647 [hep-ph]}}.

\bibitem{Okada:2019sbb}
N.~Okada and O.~Seto, ``{Inelastic extra $U(1)$ charged scalar dark matter},''
  \href{http://dx.doi.org/10.1103/PhysRevD.101.023522}{{\em Phys. Rev. D} {\bf
  101} (2020) no.~2, 023522}, \href{http://arxiv.org/abs/1908.09277}{{\tt
  arXiv:1908.09277 [hep-ph]}}.

\bibitem{Holst:2021lzm}
I.~Holst, D.~Hooper, and G.~Krnjaic, ``{Simplest and Most Predictive Model of
  Muon g-2 and Thermal Dark Matter},''
  \href{http://dx.doi.org/10.1103/PhysRevLett.128.141802}{{\em Phys. Rev.
  Lett.} {\bf 128} (2022) no.~14, 141802},
  \href{http://arxiv.org/abs/2107.09067}{{\tt arXiv:2107.09067 [hep-ph]}}.

\bibitem{Baek:2001kca}
S.~Baek, N.~G. Deshpande, X.~G. He, and P.~Ko, ``{Muon anomalous g-2 and gauged
  L(muon) - L(tau) models},''
  \href{http://dx.doi.org/10.1103/PhysRevD.64.055006}{{\em Phys. Rev. D} {\bf
  64} (2001)  055006}, \href{http://arxiv.org/abs/hep-ph/0104141}{{\tt
  arXiv:hep-ph/0104141}}.

\bibitem{Ma:2001md}
E.~Ma, D.~P. Roy, and S.~Roy, ``{Gauged L(mu) - L(tau) with large muon
  anomalous magnetic moment and the bimaximal mixing of neutrinos},''
  \href{http://dx.doi.org/10.1016/S0370-2693(01)01428-9}{{\em Phys. Lett. B}
  {\bf 525} (2002)  101--106}, \href{http://arxiv.org/abs/hep-ph/0110146}{{\tt
  arXiv:hep-ph/0110146}}.

\bibitem{Harigaya:2013twa}
K.~Harigaya, T.~Igari, M.~M. Nojiri, M.~Takeuchi, and K.~Tobe, ``{Muon g-2 and
  LHC phenomenology in the $L_\mu-L_\tau$ gauge symmetric model},''
  \href{http://dx.doi.org/10.1007/JHEP03(2014)105}{{\em JHEP} {\bf 03} (2014)
  105}, \href{http://arxiv.org/abs/1311.0870}{{\tt arXiv:1311.0870 [hep-ph]}}.

\bibitem{Altmannshofer:2016brv}
W.~Altmannshofer, C.-Y. Chen, P.~S. Bhupal~Dev, and A.~Soni, ``{Lepton flavor
  violating Z' explanation of the muon anomalous magnetic moment},''
  \href{http://dx.doi.org/10.1016/j.physletb.2016.09.046}{{\em Phys. Lett. B}
  {\bf 762} (2016)  389--398}, \href{http://arxiv.org/abs/1607.06832}{{\tt
  arXiv:1607.06832 [hep-ph]}}.

\bibitem{Escudero:2019gzq}
M.~Escudero, D.~Hooper, G.~Krnjaic, and M.~Pierre, ``{Cosmology with A Very
  Light ${L_\mu-L_\tau}$ Gauge Boson},''
  \href{http://dx.doi.org/10.1007/JHEP03(2019)071}{{\em JHEP} {\bf 03} (2019)
  071}, \href{http://arxiv.org/abs/1901.02010}{{\tt arXiv:1901.02010
  [hep-ph]}}.

\bibitem{Carpio:2021jhu}
J.~A. Carpio, K.~Murase, I.~M. Shoemaker, and Z.~Tabrizi, ``{High-energy cosmic
  neutrinos as a probe of the vector mediator scenario in light of the muon g-2
  anomaly and Hubble tension},''
  \href{http://dx.doi.org/10.1103/PhysRevD.107.103057}{{\em Phys. Rev. D} {\bf
  107} (2023) no.~10, 103057}, \href{http://arxiv.org/abs/2104.15136}{{\tt
  arXiv:2104.15136 [hep-ph]}}.

\bibitem{Araki:2021xdk}
T.~Araki, K.~Asai, K.~Honda, R.~Kasuya, J.~Sato, T.~Shimomura, and M.~J.~S.
  Yang, ``{Resolving the Hubble tension in a U(1)$_{L_\mu-L_\tau}$ model with
  the Majoron},'' \href{http://dx.doi.org/10.1093/ptep/ptab108}{{\em PTEP} {\bf
  2021} (2021) no.~10, 103B05}, \href{http://arxiv.org/abs/2103.07167}{{\tt
  arXiv:2103.07167 [hep-ph]}}.

\bibitem{Altmannshofer:2014cfa}
W.~Altmannshofer, S.~Gori, M.~Pospelov, and I.~Yavin, ``{Quark flavor
  transitions in $L_\mu-L_\tau$ models},''
  \href{http://dx.doi.org/10.1103/PhysRevD.89.095033}{{\em Phys. Rev. D} {\bf
  89} (2014)  095033}, \href{http://arxiv.org/abs/1403.1269}{{\tt
  arXiv:1403.1269 [hep-ph]}}.

\bibitem{Crivellin:2015mga}
A.~Crivellin, G.~D'Ambrosio, and J.~Heeck, ``{Explaining $h\to\mu^\pm\tau^\mp$,
  $B\to K^* \mu^+\mu^-$ and $B\to K \mu^+\mu^-/B\to K e^+e^-$ in a
  two-Higgs-doublet model with gauged $L_\mu-L_\tau$},''
  \href{http://dx.doi.org/10.1103/PhysRevLett.114.151801}{{\em Phys. Rev.
  Lett.} {\bf 114} (2015)  151801}, \href{http://arxiv.org/abs/1501.00993}{{\tt
  arXiv:1501.00993 [hep-ph]}}.

\bibitem{Altmannshofer:2016jzy}
W.~Altmannshofer, S.~Gori, S.~Profumo, and F.~S. Queiroz, ``{Explaining dark
  matter and B decay anomalies with an $L_\mu - L_\tau$ model},''
  \href{http://dx.doi.org/10.1007/JHEP12(2016)106}{{\em JHEP} {\bf 12} (2016)
  106}, \href{http://arxiv.org/abs/1609.04026}{{\tt arXiv:1609.04026
  [hep-ph]}}.

\bibitem{Chen:2017usq}
C.-H. Chen and T.~Nomura, ``{Penguin $b \to s\ell'^+ \ell'^-$ and $B$-meson
  anomalies in a gauged ${L_\mu -L_\tau}$},''
  \href{http://dx.doi.org/10.1016/j.physletb.2017.12.062}{{\em Phys. Lett. B}
  {\bf 777} (2018)  420--427}, \href{http://arxiv.org/abs/1707.03249}{{\tt
  arXiv:1707.03249 [hep-ph]}}.

\bibitem{Baek:2017sew}
S.~Baek, ``{Dark matter contribution to $b\to s \mu^+ \mu^-$ anomaly in local
  $U(1)_{L_\mu-L_\tau}$ model},''
  \href{http://dx.doi.org/10.1016/j.physletb.2018.04.012}{{\em Phys. Lett. B}
  {\bf 781} (2018)  376--382}, \href{http://arxiv.org/abs/1707.04573}{{\tt
  arXiv:1707.04573 [hep-ph]}}.

\bibitem{Altmannshofer:2023uci}
W.~Altmannshofer, S.~A. Gadam, and S.~Profumo, ``{Probing new physics with
  \ensuremath{\mu}+\ensuremath{\mu}-\textrightarrow{}bs at a muon collider},''
  \href{http://dx.doi.org/10.1103/PhysRevD.108.115033}{{\em Phys. Rev. D} {\bf
  108} (2023) no.~11, 115033}, \href{http://arxiv.org/abs/2306.15017}{{\tt
  arXiv:2306.15017 [hep-ph]}}.

\bibitem{LHCb:2022qnv}
{\bf LHCb} Collaboration, R.~Aaij {\em et al.}, ``{Test of lepton universality
  in $b \rightarrow s \ell^+ \ell^-$ decays},''
  \href{http://dx.doi.org/10.1103/PhysRevLett.131.051803}{{\em Phys. Rev.
  Lett.} {\bf 131} (2023) no.~5, 051803},
  \href{http://arxiv.org/abs/2212.09152}{{\tt arXiv:2212.09152 [hep-ex]}}.

\bibitem{LHCb:2022vje}
{\bf LHCb} Collaboration, R.~Aaij {\em et al.}, ``{Measurement of lepton
  universality parameters in $B^+\to K^+\ell^+\ell^-$ and $B^0\to
  K^{*0}\ell^+\ell^-$ decays},''
  \href{http://dx.doi.org/10.1103/PhysRevD.108.032002}{{\em Phys. Rev. D} {\bf
  108} (2023) no.~3, 032002}, \href{http://arxiv.org/abs/2212.09153}{{\tt
  arXiv:2212.09153 [hep-ex]}}.

\bibitem{Amaral:2021rzw}
D.~W.~P. Amaral, D.~G. Cerdeno, A.~Cheek, and P.~Foldenauer, ``{Confirming
  $U(1)_{L_\mu -L_{\tau }}$ as a solution for $(g-2)_\mu $ with neutrinos},''
  \href{http://dx.doi.org/10.1140/epjc/s10052-021-09670-z}{{\em Eur. Phys. J.
  C} {\bf 81} (2021) no.~10, 861}, \href{http://arxiv.org/abs/2104.03297}{{\tt
  arXiv:2104.03297 [hep-ph]}}.

\bibitem{Kaneta:2016uyt}
Y.~Kaneta and T.~Shimomura, ``{On the possibility of a search for the $L_\mu -
  L_\tau$ gauge boson at Belle-II and neutrino beam experiments},''
  \href{http://dx.doi.org/10.1093/ptep/ptx050}{{\em PTEP} {\bf 2017} (2017)
  no.~5, 053B04}, \href{http://arxiv.org/abs/1701.00156}{{\tt arXiv:1701.00156
  [hep-ph]}}.

\bibitem{Amaral:2020tga}
D.~W. P.~d. Amaral, D.~G. Cerdeno, P.~Foldenauer, and E.~Reid, ``{Solar
  neutrino probes of the muon anomalous magnetic moment in the gauged $
  \mathrm{U}{(1)}_{L_{\mu }-{L}_{\tau }} $},''
  \href{http://dx.doi.org/10.1007/JHEP12(2020)155}{{\em JHEP} {\bf 12} (2020)
  155}, \href{http://arxiv.org/abs/2006.11225}{{\tt arXiv:2006.11225
  [hep-ph]}}.

\bibitem{Kamada:2015era}
A.~Kamada and H.-B. Yu, ``{Coherent Propagation of PeV Neutrinos and the Dip in
  the Neutrino Spectrum at IceCube},''
  \href{http://dx.doi.org/10.1103/PhysRevD.92.113004}{{\em Phys. Rev. D} {\bf
  92} (2015) no.~11, 113004}, \href{http://arxiv.org/abs/1504.00711}{{\tt
  arXiv:1504.00711 [hep-ph]}}.

\bibitem{Kamada:2018zxi}
A.~Kamada, K.~Kaneta, K.~Yanagi, and H.-B. Yu, ``{Self-interacting dark matter
  and muon $g-2$ in a gauged U$(1)_{L_{\mu} - L_{\tau}}$ model},''
  \href{http://dx.doi.org/10.1007/JHEP06(2018)117}{{\em JHEP} {\bf 06} (2018)
  117}, \href{http://arxiv.org/abs/1805.00651}{{\tt arXiv:1805.00651
  [hep-ph]}}.

\bibitem{Croon:2020lrf}
D.~Croon, G.~Elor, R.~K. Leane, and S.~D. McDermott, ``{Supernova Muons: New
  Constraints on $Z$' Bosons, Axions and ALPs},''
  \href{http://dx.doi.org/10.1007/JHEP01(2021)107}{{\em JHEP} {\bf 01} (2021)
  107}, \href{http://arxiv.org/abs/2006.13942}{{\tt arXiv:2006.13942
  [hep-ph]}}.

\bibitem{Cerdeno:2023kqo}
D.~G. Cerde\~no, M.~Cerme\~no, and Y.~Farzan, ``{Constraints from the duration
  of supernova neutrino burst on on-shell light gauge boson production by
  neutrinos},'' \href{http://dx.doi.org/10.1103/PhysRevD.107.123012}{{\em Phys.
  Rev. D} {\bf 107} (2023) no.~12, 123012},
  \href{http://arxiv.org/abs/2301.00661}{{\tt arXiv:2301.00661 [hep-ph]}}.

\bibitem{Akita:2023iwq}
K.~Akita, S.~H. Im, M.~Masud, and S.~Yun, ``{Limits on heavy neutral leptons,
  $Z'$ bosons and majorons from high-energy supernova neutrinos},''
  \href{http://arxiv.org/abs/2312.13627}{{\tt arXiv:2312.13627 [hep-ph]}}.

\bibitem{Feynman:1949zz}
R.~P. Feynman, N.~Metropolis, and E.~Teller, ``{Equations of State of Elements
  Based on the Generalized Fermi-Thomas Theory},''
  \href{http://dx.doi.org/10.1103/PhysRev.75.1561}{{\em Phys. Rev.} {\bf 75}
  (1949)  1561--1573}.

\bibitem{Salpeter:1961zz}
E.~E. Salpeter, ``{Energy and Pressure of a Zero-Temperature Plasma},''
  \href{http://dx.doi.org/10.1086/147194}{{\em Astrophys. J.} {\bf 134} (1961)
  669--682}.

\bibitem{Rotondo:2011zz}
M.~Rotondo, J.~A. Rueda, R.~Ruffini, and S.-S. Xue, ``{The Relativistic
  Feynman-Metropolis-Teller theory for white dwarfs in general relativity},''
  \href{http://dx.doi.org/10.1103/PhysRevD.84.084007}{{\em Phys. Rev. D} {\bf
  84} (2011)  084007}, \href{http://arxiv.org/abs/1012.0154}{{\tt
  arXiv:1012.0154 [astro-ph.SR]}}.

\bibitem{mathew2017general}
A.~Mathew and M.~K. Nandy, ``General relativistic calculations for white
  dwarfs,'' \href{http://dx.doi.org/10.1088/1674-4527/17/6/61}{{\em Research in
  Astronomy and Astrophysics} {\bf 17} (2017) no.~6, 061}.

\bibitem{Fantoni:2017mfs}
R.~Fantoni, ``{White-dwarf equation of state and structure: the effect of
  temperature},'' \href{http://dx.doi.org/10.1088/1742-5468/aa9339}{{\em J.
  Stat. Mech.} {\bf 1711} (2017) no.~11, 113101},
  \href{http://arxiv.org/abs/1709.06064}{{\tt arXiv:1709.06064 [astro-ph.SR]}}.

\bibitem{Winget:2003xf}
D.~E. Winget, D.~J. Sullivan, T.~S. Metcalfe, S.~D. Kawaler, and M.~H.
  Montgomery, ``{A strong test of electro-weak theory using pulsating db white
  dwarf stars as plasmon neutrino detectors},''
  \href{http://dx.doi.org/10.1086/382591}{{\em Astrophys. J. Lett.} {\bf 602}
  (2004)  L109--L112}, \href{http://arxiv.org/abs/astro-ph/0312303}{{\tt
  arXiv:astro-ph/0312303}}.

\bibitem{Kantor:2007kf}
E.~M. Kantor and M.~E. Gusakov, ``{The neutrino emission due to plasmon decay
  and neutrino luminosity of white dwarfs},''
  \href{http://dx.doi.org/10.1111/j.1365-2966.2007.12342.x}{{\em Mon. Not. Roy.
  Astron. Soc.} {\bf 381} (2007)  1702},
  \href{http://arxiv.org/abs/0708.2093}{{\tt arXiv:0708.2093 [astro-ph]}}.

\bibitem{Landstreet:1967zz}
J.~D. Landstreet, ``{Synchrotron Radiation of Neutrinos and Its Astrophysical
  Significance},'' \href{http://dx.doi.org/10.1103/PhysRev.153.1372}{{\em Phys.
  Rev.} {\bf 153} (1967)  1372--1377}.

\bibitem{chaudhuri1970neutrino}
P.~R. {Chaudhuri}, ``{Neutrino Synchrotron Radiation. I: Application to White
  Dwarfs},'' \href{http://dx.doi.org/10.1007/BF00651335}{{\em Astrophysics and
  Space Science} {\bf 8} (1970) no.~3, 432--447}.

\bibitem{ca00110y}
V.~Canuto, C.~Chiuderi, and C.~K. Chou, ``Plasmon neutrinos emission in a
  strong magnetic field. i: Transverse plasmons,''
  \href{http://dx.doi.org/10.1007/BF00653279}{{\em Astrophys. Space Sci.} {\bf
  7} (1970)  407--415}.

\bibitem{ca01110v}
V.~Canuto, C.~Chiuderi, and C.~K. Chou, ``Plasmon neutrinos emission in a
  strong magnetic field. ii: Longitudinal plasmons,''
  \href{http://dx.doi.org/10.1007/BF00649583}{{\em Astrophys. Space Sci.} {\bf
  7} (1970)  453--460}.

\bibitem{galtsov1972photoneutrino}
D.~Galtsov and N.~Nikitina, ``Photoneutrino processes in a strong field,'' {\em
  Sov. Phys. JETP} {\bf 35} (1972)  1047.

\bibitem{DeRaad:1976kd}
L.~L. DeRaad, Jr., K.~A. Milton, and N.~D. Hari~Dass, ``{Photon Decay Into
  Neutrinos in a Strong Magnetic Field},''
  \href{http://dx.doi.org/10.1103/PhysRevD.14.3326}{{\em Phys. Rev. D} {\bf 14}
  (1976)  3326}.

\bibitem{skobelev1976reaction}
V.~Skobelev, ``Reaction whereby a photon decays into a neutrino-antineutrino
  pair and a neutrino decays into a photon-neutrino pair in a strong magnetic
  field,'' {\em Zhurnal Eksperimentalnoi i Teoreticheskoi Fiziki} {\bf 71}
  (1976)  1263--1267.

\bibitem{Yakovlev1981}
D.~G. Yakovlev and R.~Tschaepe, ``Synchrotron neutrino-pair radiation in
  neutron stars,''
  \href{http://dx.doi.org/https://doi.org/10.1002/asna.19813020401}{{\em
  Astronomische Nachrichten} {\bf 302} (1981) no.~4, 167--176}.

\bibitem{Kaminker:1992su}
A.~D. Kaminker, K.~P. Levenfish, D.~G. Yakovlev, P.~Amsterdamski, and
  P.~Haensel, ``{Neutrino emissivity from e- synchrotron and e- e+ annihilation
  processes in a strong magnetic field: General formalism and nonrelativistic
  limit},'' \href{http://dx.doi.org/10.1103/PhysRevD.46.3256}{{\em Phys. Rev.
  D} {\bf 46} (1992)  3256--3264}.

\bibitem{Kennett:1999jh}
M.~P. Kennett and D.~B. Melrose, ``{Neutrino emission via the plasma process in
  a magnetized plasma},''
  \href{http://dx.doi.org/10.1103/PhysRevD.58.093011}{{\em Phys. Rev. D} {\bf
  58} (1998)  093011}, \href{http://arxiv.org/abs/astro-ph/9901156}{{\tt
  arXiv:astro-ph/9901156}}.

\bibitem{Bhattacharyya:2005tf}
I.~Bhattacharyya, ``{Neutrino synchrotron radiation in electro-weak
  interaction},''
  \href{http://dx.doi.org/10.1016/j.astropartphys.2005.06.003}{{\em Astropart.
  Phys.} {\bf 24} (2005)  100--106}.

\bibitem{Drewes:2021fjx}
M.~Drewes, J.~McDonald, L.~Sablon, and E.~Vitagliano, ``{Neutrino Emissivities
  as a Probe of the Internal Magnetic Fields of White Dwarfs},''
  \href{http://dx.doi.org/10.3847/1538-4357/ac7874}{{\em Astrophys. J.} {\bf
  934} (2022)  99}, \href{http://arxiv.org/abs/2109.06158}{{\tt
  arXiv:2109.06158 [astro-ph.SR]}}.

\bibitem{Bauer:2018onh}
M.~Bauer, P.~Foldenauer, and J.~Jaeckel, ``{Hunting All the Hidden Photons},''
  \href{http://dx.doi.org/10.1007/JHEP07(2018)094}{{\em JHEP} {\bf 07} (2018)
  094}, \href{http://arxiv.org/abs/1803.05466}{{\tt arXiv:1803.05466
  [hep-ph]}}.

\bibitem{Dreiner:2013tja}
H.~K. Dreiner, J.-F. Fortin, J.~Isern, and L.~Ubaldi, ``{White Dwarfs constrain
  Dark Forces},'' \href{http://dx.doi.org/10.1103/PhysRevD.88.043517}{{\em
  Phys. Rev. D} {\bf 88} (2013)  043517},
  \href{http://arxiv.org/abs/1303.7232}{{\tt arXiv:1303.7232 [hep-ph]}}.

\bibitem{Planck:2018vyg}
{\bf Planck} Collaboration, N.~Aghanim {\em et al.}, ``{Planck 2018 results.
  VI. Cosmological parameters},''
  \href{http://dx.doi.org/10.1051/0004-6361/201833910}{{\em Astron. Astrophys.}
  {\bf 641} (2020)  A6}, \href{http://arxiv.org/abs/1807.06209}{{\tt
  arXiv:1807.06209 [astro-ph.CO]}}. [Erratum: Astron.Astrophys. 652, C4
  (2021)].

\bibitem{Riess:2019cxk}
A.~G. Riess, S.~Casertano, W.~Yuan, L.~M. Macri, and D.~Scolnic, ``{Large
  Magellanic Cloud Cepheid Standards Provide a 1\% Foundation for the
  Determination of the Hubble Constant and Stronger Evidence for Physics beyond
  $\Lambda$CDM},'' \href{http://dx.doi.org/10.3847/1538-4357/ab1422}{{\em
  Astrophys. J.} {\bf 876} (2019) no.~1, 85},
  \href{http://arxiv.org/abs/1903.07603}{{\tt arXiv:1903.07603 [astro-ph.CO]}}.

\bibitem{Blinov:2019gcj}
N.~Blinov, K.~J. Kelly, G.~Z. Krnjaic, and S.~D. McDermott, ``{Constraining the
  Self-Interacting Neutrino Interpretation of the Hubble Tension},''
  \href{http://dx.doi.org/10.1103/PhysRevLett.123.191102}{{\em Phys. Rev.
  Lett.} {\bf 123} (2019) no.~19, 191102},
  \href{http://arxiv.org/abs/1905.02727}{{\tt arXiv:1905.02727 [astro-ph.CO]}}.

\bibitem{Bauer:2022nwt}
M.~Bauer and P.~Foldenauer, ``{Consistent Theory of Kinetic Mixing and the
  Higgs Low-Energy Theorem},''
  \href{http://dx.doi.org/10.1103/PhysRevLett.129.171801}{{\em Phys. Rev.
  Lett.} {\bf 129} (2022) no.~17, 171801},
  \href{http://arxiv.org/abs/2207.00023}{{\tt arXiv:2207.00023 [hep-ph]}}.

\bibitem{Lynch:2001zs}
K.~R. Lynch, ``{A Note on one loop electroweak contributions to g-2: A
  Companion to BUHEP-01-16},'' \href{http://arxiv.org/abs/hep-ph/0108081}{{\tt
  arXiv:hep-ph/0108081}}.

\bibitem{Pospelov:2008zw}
M.~Pospelov, ``{Secluded U(1) below the weak scale},''
  \href{http://dx.doi.org/10.1103/PhysRevD.80.095002}{{\em Phys. Rev. D} {\bf
  80} (2009)  095002}, \href{http://arxiv.org/abs/0811.1030}{{\tt
  arXiv:0811.1030 [hep-ph]}}.

\bibitem{Aoyama:2020ynm}
T.~Aoyama {\em et al.}, ``{The anomalous magnetic moment of the muon in the
  Standard Model},''
  \href{http://dx.doi.org/10.1016/j.physrep.2020.07.006}{{\em Phys. Rept.} {\bf
  887} (2020)  1--166}, \href{http://arxiv.org/abs/2006.04822}{{\tt
  arXiv:2006.04822 [hep-ph]}}.

\bibitem{Muong-2:2023cdq}
{\bf Muon g-2} Collaboration, D.~P. Aguillard {\em et al.}, ``{Measurement of
  the Positive Muon Anomalous Magnetic Moment to 0.20~ppm},''
  \href{http://dx.doi.org/10.1103/PhysRevLett.131.161802}{{\em Phys. Rev.
  Lett.} {\bf 131} (2023) no.~16, 161802},
  \href{http://arxiv.org/abs/2308.06230}{{\tt arXiv:2308.06230 [hep-ex]}}.

\bibitem{Muong-2:2021ojo}
{\bf Muon g-2} Collaboration, B.~Abi {\em et al.}, ``{Measurement of the
  Positive Muon Anomalous Magnetic Moment to 0.46 ppm},''
  \href{http://dx.doi.org/10.1103/PhysRevLett.126.141801}{{\em Phys. Rev.
  Lett.} {\bf 126} (2021) no.~14, 141801},
  \href{http://arxiv.org/abs/2104.03281}{{\tt arXiv:2104.03281 [hep-ex]}}.

\bibitem{Muong-2:2006rrc}
{\bf Muon g-2} Collaboration, G.~W. Bennett {\em et al.}, ``{Final Report of
  the Muon E821 Anomalous Magnetic Moment Measurement at BNL},''
  \href{http://dx.doi.org/10.1103/PhysRevD.73.072003}{{\em Phys. Rev. D} {\bf
  73} (2006)  072003}, \href{http://arxiv.org/abs/hep-ex/0602035}{{\tt
  arXiv:hep-ex/0602035}}.

\bibitem{Borsanyi:2020mff}
S.~Borsanyi {\em et al.}, ``{Leading hadronic contribution to the muon magnetic
  moment from lattice QCD},''
  \href{http://dx.doi.org/10.1038/s41586-021-03418-1}{{\em Nature} {\bf 593}
  (2021) no.~7857, 51--55}, \href{http://arxiv.org/abs/2002.12347}{{\tt
  arXiv:2002.12347 [hep-lat]}}.

\bibitem{Crivellin:2020zul}
A.~Crivellin, M.~Hoferichter, C.~A. Manzari, and M.~Montull, ``{Hadronic Vacuum
  Polarization: $(g-2)_\mu$ versus Global Electroweak Fits},''
  \href{http://dx.doi.org/10.1103/PhysRevLett.125.091801}{{\em Phys. Rev.
  Lett.} {\bf 125} (2020) no.~9, 091801},
  \href{http://arxiv.org/abs/2003.04886}{{\tt arXiv:2003.04886 [hep-ph]}}.

\bibitem{Zink:2023szx}
J.~H. Zink and M.~E. Ramirez-Quezada, ``{Exploring the dark sectors via the
  cooling of white dwarfs},''
  \href{http://dx.doi.org/10.1103/PhysRevD.108.043014}{{\em Phys. Rev. D} {\bf
  108} (2023) no.~4, 043014}, \href{http://arxiv.org/abs/2306.00517}{{\tt
  arXiv:2306.00517 [hep-ph]}}.

\bibitem{Braaten:1993jw}
E.~Braaten and D.~Segel, ``{Neutrino energy loss from the plasma process at all
  temperatures and densities},''
  \href{http://dx.doi.org/10.1103/PhysRevD.48.1478}{{\em Phys. Rev. D} {\bf 48}
  (1993)  1478--1491}, \href{http://arxiv.org/abs/hep-ph/9302213}{{\tt
  arXiv:hep-ph/9302213}}.

\bibitem{shapiro1983physics}
S.~L. {Shapiro} and S.~A. {Teukolsky}, ``{Black holes, white dwarfs and neutron
  stars. The physics of compact objects},''
  \href{http://dx.doi.org/10.1002/9783527617661}{{\em Wiley, New York} {\bf
  19832} (1983)  119--123}.

\bibitem{Breit:1936zzb}
G.~Breit and E.~Wigner, ``{Capture of Slow Neutrons},''
  \href{http://dx.doi.org/10.1103/PhysRev.49.519}{{\em Phys. Rev.} {\bf 49}
  (1936)  519--531}.

\bibitem{Weldon:1983jn}
H.~A. Weldon, ``{Simple Rules for Discontinuities in Finite Temperature Field
  Theory},'' \href{http://dx.doi.org/10.1103/PhysRevD.28.2007}{{\em Phys. Rev.
  D} {\bf 28} (1983)  2007}.

\bibitem{Lepage:2020tgj}
G.~P. Lepage, ``{Adaptive multidimensional integration: VEGAS enhanced},''
  \href{http://dx.doi.org/10.1016/j.jcp.2021.110386}{{\em J. Comput. Phys.}
  {\bf 439} (2021)  110386}, \href{http://arxiv.org/abs/2009.05112}{{\tt
  arXiv:2009.05112 [physics.comp-ph]}}.

\bibitem{Hansen:2015lqa}
B.~M.~S. Hansen, H.~Richer, J.~Kalirai, R.~Goldsbury, S.~Frewen, and J.~Heyl,
  ``{Constraining Neutrino Cooling using the Hot White Dwarf Luminosity
  Function in the Globular Cluster 47 Tucanae},''
  \href{http://dx.doi.org/10.1088/0004-637X/809/2/141}{{\em Astrophys. J.} {\bf
  809} (2015) no.~2, 141}, \href{http://arxiv.org/abs/1507.05665}{{\tt
  arXiv:1507.05665 [astro-ph.SR]}}.

\bibitem{Andreev:2024sgn}
Y.~M. Andreev {\em et al.}, ``{Exploration of the Muon $g-2$ and Light Dark
  Matter explanations in NA64 with the CERN SPS high energy muon beam},''
  \href{http://arxiv.org/abs/2401.01708}{{\tt arXiv:2401.01708 [hep-ex]}}.

\bibitem{Bellini:2011rx}
G.~Bellini {\em et al.}, ``{Precision measurement of the 7Be solar neutrino
  interaction rate in Borexino},''
  \href{http://dx.doi.org/10.1103/PhysRevLett.107.141302}{{\em Phys. Rev.
  Lett.} {\bf 107} (2011)  141302}, \href{http://arxiv.org/abs/1104.1816}{{\tt
  arXiv:1104.1816 [hep-ex]}}.

\bibitem{Borexino:2017rsf}
{\bf Borexino} Collaboration, M.~Agostini {\em et al.}, ``{First Simultaneous
  Precision Spectroscopy of $pp$, $^7$Be, and $pep$ Solar Neutrinos with
  Borexino Phase-II},''
  \href{http://dx.doi.org/10.1103/PhysRevD.100.082004}{{\em Phys. Rev. D} {\bf
  100} (2019) no.~8, 082004}, \href{http://arxiv.org/abs/1707.09279}{{\tt
  arXiv:1707.09279 [hep-ex]}}.

\bibitem{BaBar:2016sci}
{\bf BaBar} Collaboration, J.~P. Lees {\em et al.}, ``{Search for a muonic dark
  force at BABAR},'' \href{http://dx.doi.org/10.1103/PhysRevD.94.011102}{{\em
  Phys. Rev. D} {\bf 94} (2016) no.~1, 011102},
  \href{http://arxiv.org/abs/1606.03501}{{\tt arXiv:1606.03501 [hep-ex]}}.

\bibitem{COHERENT:2015mry}
{\bf COHERENT} Collaboration, D.~Akimov {\em et al.}, ``{The COHERENT
  Experiment at the Spallation Neutron Source},''
  \href{http://arxiv.org/abs/1509.08702}{{\tt arXiv:1509.08702
  [physics.ins-det]}}.

\bibitem{COHERENT:2017ipa}
{\bf COHERENT} Collaboration, D.~Akimov {\em et al.}, ``{Observation of
  Coherent Elastic Neutrino-Nucleus Scattering},''
  \href{http://dx.doi.org/10.1126/science.aao0990}{{\em Science} {\bf 357}
  (2017) no.~6356, 1123--1126}, \href{http://arxiv.org/abs/1708.01294}{{\tt
  arXiv:1708.01294 [nucl-ex]}}.

\bibitem{Altmannshofer:2014pba}
W.~Altmannshofer, S.~Gori, M.~Pospelov, and I.~Yavin, ``{Neutrino Trident
  Production: A Powerful Probe of New Physics with Neutrino Beams},''
  \href{http://dx.doi.org/10.1103/PhysRevLett.113.091801}{{\em Phys. Rev.
  Lett.} {\bf 113} (2014)  091801}, \href{http://arxiv.org/abs/1406.2332}{{\tt
  arXiv:1406.2332 [hep-ph]}}.

\bibitem{CHARM-II:1990dvf}
{\bf CHARM-II} Collaboration, D.~Geiregat {\em et al.}, ``{First observation of
  neutrino trident production},''
  \href{http://dx.doi.org/10.1016/0370-2693(90)90146-W}{{\em Phys. Lett. B}
  {\bf 245} (1990)  271--275}.

\bibitem{Krnjaic:2019rsv}
G.~Krnjaic, G.~Marques-Tavares, D.~Redigolo, and K.~Tobioka, ``{Probing
  Muonphilic Force Carriers and Dark Matter at Kaon Factories},''
  \href{http://dx.doi.org/10.1103/PhysRevLett.124.041802}{{\em Phys. Rev.
  Lett.} {\bf 124} (2020) no.~4, 041802},
  \href{http://arxiv.org/abs/1902.07715}{{\tt arXiv:1902.07715 [hep-ph]}}.

\bibitem{10.1117/12.926198}
P.~C{\^o}t\'e {\em et al.},
  \href{http://dx.doi.org/10.1117/12.926198}{``{CASTOR: the Cosmological
  Advanced Survey Telescope for Optical and Ultraviolet Research},''} in {\em
  Space Telescopes and Instrumentation 2012: Optical, Infrared, and Millimeter
  Wave}, M.~C. Clampin, G.~G. Fazio, H.~A. MacEwen, and J.~M.~O. Jr., eds.,
  vol.~8442, p.~844215, International Society for Optics and Photonics.
\newblock SPIE, 2012.

\bibitem{Fantin_2020}
N.~J. Fantin, P.~Côté, and A.~W. McConnachie, ``{White Dwarfs in the Era of
  the LSST and Its Synergies with Space-based Missions},''
  \href{http://dx.doi.org/10.3847/1538-4357/aba270}{{\em The Astrophysical
  Journal} {\bf 900} (2020) no.~2, 139},
  \href{http://arxiv.org/abs/2007.01312}{{\tt arXiv:2007.01312 [astro-ph.GA]}}.

\bibitem{castor19}
P.~Côté {\em et al.}, ``Castor: A flagship canadian space telescope id of
  associated expression of interest topic area of white paper executive summary
  of white paper.'' Zenodo, 10, 2019.
\newblock \url{https://doi.org/10.5281/zenodo.3758463}.

\bibitem{peter_lepage_2024_10783443}
P.~Lepage, ``gplepage/vegas: vegas version 6.0.'' Zenodo, Mar., 2024.
\newblock \url{https://doi.org/10.5281/zenodo.10783443}.

\end{thebibliography}\endgroup

\end{document}